\documentclass[journal,10pt]{IEEEtran}
\makeatletter
\def\endthebibliography{%
    \def\@noitemerr{\@latex@warning{Empty `thebibliography' environment}}%
    \endlist
}
\makeatother

\usepackage{cite}

\usepackage[pdftex]{graphicx}
\usepackage[caption=false,font=footnotesize]{subfig}

\usepackage{amsmath}
\usepackage{mathtools, cuted}
\usepackage{amssymb}
\usepackage{bm}
\usepackage{mathrsfs}
\usepackage{breqn}

\usepackage{pifont}
\usepackage{xcolor}
\usepackage{url}
\usepackage{lettrine}
\usepackage{lipsum}
\usepackage{siunitx}
\usepackage{soul}
\usepackage{array}
\usepackage[inline]{enumitem}
\usepackage{epsfig}
\usepackage{filecontents}
\usepackage{booktabs}

\usepackage{algpseudocode}
\usepackage{algorithm, tabularx}
\usepackage{multirow}
\newcolumntype{L}[1]{>{\raggedright\let\newline\\\arraybackslash\hspace{0pt}}m{#1}}
\newcolumntype{C}[1]{>{\centering\let\newline\\\arraybackslash\hspace{0pt}}m{#1}}
\newcolumntype{R}[1]{>{\raggedleft\let\newline\\\arraybackslash\hspace{0pt}}m{#1}}
\newlength{\maxwidth}

\makeatletter
\newcommand{\multiline}[1]{%
	\begin{tabularx}{\dimexpr\linewidth-\ALG@thistlm}[t]{@{}X@{}}
		#1
	\end{tabularx}
}
\makeatother
\algdef{SE}[SUBALG]{Indent}{EndIndent}{}{{\algorithmicend\ }}
\algtext*{Indent}
\algtext*{EndIndent}

\usepackage{amsthm}

\theoremstyle{remark}

\newtheorem{remark}{Remark}

\begin{document}

\title{RIS-aided ISAC with $K$-Rydberg Atomic Receivers}

\author{Hong-Bae Jeon,~\IEEEmembership{Member,~IEEE,} and Chan-Byoung Chae,~\IEEEmembership{Fellow,~IEEE}%
\thanks{This work was supported by Hankuk University of Foreign Studies Research Fund of 2026. \textit{(Corresponding: Chan-Byoung Chae.)}}%
\thanks{H.-B. Jeon is with the Department of Information Communications Engineering, Hankuk University of Foreign Studies, Yong-in, 17035, Korea (e-mail: hongbae08@hufs.ac.kr).}%
\thanks{C.-B. Chae is with the School of Integrated Technology, Yonsei University, Seoul 03722, Korea (e-mail: cbchae@yonsei.ac.kr).}%
}

\maketitle

\begin{abstract}
In this paper, we investigate a reconfigurable intelligent surface (RIS)-assisted integrated sensing and communications (ISAC) framework equipped with multiple Rydberg atomic receiver (RAR)-aided users. By leveraging the reference-assisted reception mechanism of RARs, we develop a unified signal model that jointly captures downlink multi-user communication with RARs and monostatic radar sensing. To explicitly balance communication performance and sensing accuracy, we formulate a Cram\'er-Rao bound (CRB)-constrained utility maximization problem. To address these challenges, we propose a joint optimization framework that combines fractional programming (FP), majorization-minimization (MM), and the alternating direction method of multipliers (ADMM). Simulation results demonstrate that the proposed framework consistently outperforms the conventional approach over a wide range of system environments, thereby highlighting the importance of the proposed framework in unlocking the potential of RARs for 6G.
\end{abstract}


\IEEEpeerreviewmaketitle

\section{Introduction}
\label{sec:intro}

\IEEEPARstart{F}{or} decades, wireless communication and radar sensing have evolved independently, driven by fundamentally different infrastructures and design philosophies. The advent of integrated sensing and communication (ISAC) overturns this long-standing separation by redefining the wireless stack and emerging as a foundational technology poised to shape sixth-generation (6G) wireless networks~\cite{isactut,isacjsac}. By exploiting the intrinsic commonalities between communication and sensing systems, ISAC moves beyond mere coexistence toward true joint co-design. As a result, ISAC is expected to become a core component of 6G infrastructures, enabling ubiquitous high-rate connectivity alongside precise environmental awareness~\cite{isacmag}.

Accordingly, extensive efforts from both academia and industry have focused on developing and validating practical ISAC architectures and deployment strategies~\cite{isaccsm}. Despite significant performance gains in communication and sensing, ISAC systems remain vulnerable to harsh propagation conditions~\cite{risisacoj}. In this context, reconfigurable intelligent surface (RIS) has emerged as a promising solution for reshaping wireless propagation environments, offering coverage enhancement and passive beamforming gains via programmable low-cost reflective elements~\cite{LingRIS,RIST}. As 6G networks increasingly integrate ISAC with programmable environments, joint RIS-ISAC frameworks provide a compelling pathway toward energy- and spectrum-efficient environmental control~\cite{risisacmag22}.


Many studies on RIS-ISAC have focused on enhancing radar detectability while maintaining acceptable communication performance. In~\cite{Luo_TVT23_RIS_ISAC}, transmit beamforming and RIS phase were jointly optimized to maximize radar signal-to-noise ratio (SNR) under communication quality-of-service (QoS) constraints. As security vulnerabilities in ISAC were subsequently identified, research expanded toward secure RIS-ISAC. Specifically,~\cite{TWC25_MovableAntenna_RIS_ISAC} introduced movable antenna (MA) technology to strengthen secure transmission, while~\cite{Chu_TVT24_Secure_RIS_ISAC} improved radar output SNR with secrecy protection via alternating optimization (AO). Concurrently, efforts broadened RIS-ISAC applicability to low-cost and distributed networks. In low-power Internet-of-Things (IoT) backscatter scenarios,~\cite{IoT23_RIS_Backscatter_ISAC} proposed a fair sensing-communication design that jointly maximizes sensing signal-to-interference-and-noise ratio (SINR) while mitigating multi-user interference. As ISAC networks further scaled spatially, cooperative multi-cell RIS-ISAC architectures were proposed in~\cite{TWC24_Cooperative_RIS_ISAC}, where multi-target sensing and multi-user communication were jointly enhanced through base-station (BS) cooperation, increasing the degree-of-freedom (DoF) of the system.

Recent efforts have also focused on RIS-ISAC designs that jointly incorporate sensing-accuracy objectives or constraints within the optimization problem. In~\cite{Liu_RIS_ISAC_CRB}, BS beamforming and RIS phase were jointly optimized for ISAC under radar-SNR or Cram\`er-Rao bound (CRB)-driven sensing constraints, demonstrating that RIS particularly benefits sensing accuracy. Furthermore,~\cite{IoT24_CRB_mmWave_RIS_ISAC} incorporated CRB minimization into hybrid beamforming, marking a conceptual transition from traditional SNR-centric sensing design toward estimation-theoretic ISAC optimization. 
These RIS-ISAC studies collectively reflect a progressive evolution, marking a clear trajectory toward practical RIS-ISAC realization in 6G networks.

Whereas RIS reshapes the wireless medium to support ISAC functionality, recent advances now shift attention to fundamentally transforming the receiver architecture. Meanwhile, Rydberg atomic receivers (RARs) is rapidly progressing at the intersection of quantum sensing and wireless communications~\cite{atomicmag}. Herein, the Rydberg atoms are highly excited quantum states of atoms that exhibit a strong interaction with incident electromagnetic (EM) waves, and when exposed to such fields, these atoms undergo electron transitions between their resonant energy levels~\cite{qira, Precoding_atomicMIMO}. RARs exploit these quantum behaviors, including the electromagnetically induced transparency (EIT) and Autler-Townes (AT) splitting, to sense these transitions and thereby recover the transmitted information~\cite{tqe}. Unlike classical radio-frequency (RF) receivers, RARs inherently circumvent the thermal-noise bottleneck of conventional metal antennas, since the atom-field interaction itself does not generate thermal noise~\cite{atomicjsac, Harnessing_RAR_Tutorial}. In addition, the quantum shot noise associated with probing the Rydberg quantum states is typically several orders of magnitude lower than the thermal noise floor~\cite{quanmobi}, which guarantees the sensitivity of the standard quantum limit (e.g., on the order of $\mathrm{nVcm^{-1}Hz^{-1/2}}$~\cite{qsens}). These attributes position RARs as a compelling technology for unlocking wireless reception under extremely weak EM fields such as satellite links or space-air-ground channels, where detecting weak signals is crucial~\cite{antsp}.

Early studies on RARs focused on verifying whether RF signals could be demodulated through quantum sensing mechanisms. In~\cite{atomicjsac}, the authors demonstrated recovery of amplitude- and frequency-modulated (AM/FM) signals and later extended this capability to the multi-user regime by formulating atomic single/multiple-input-multiple-output (SIMO/MIMO) detection as a biased phase-retrieval problem, along with an expectation-maximization Gerchberg-Saxton (EM-GS) algorithm for jointly decoding multi-user symbols. On the transmission side,~\cite{Precoding_atomicMIMO} introduced an atomic-MIMO communication model and highlighted its key distinction from conventional RF MIMO, namely, a nonlinear magnitude-only input-output relation, while proposing an in-phase/quadrature (IQ)-aware precoding strategy that theoretically achieves the atomic-MIMO capacity limit.

In parallel, quantum-enhanced sensing techniques have emerged. Specifically,~\cite{qmusic} proposed Quantum multiple signal classification (MUSIC) for multi-user angle-of-arrival (AoA) estimation in RARs, while\cite{wsat} extended RAR sensing to a multi-band regime by estimating AoA across different carrier frequencies. A notable advance in~\cite{TCOM25_Single_RAR_AoA} demonstrated that even a single RAR can infer AoA by exploiting inner-vapor interference within the atomic medium.

More recently, research has shifted toward system-level deployment. For instance,~\cite{RAR_MU_MIMO_Uplink} showed that a single RAR front-end can jointly demodulate spatially multiplexed uplink signals via quantum-optical processing, validating multi-user connectivity. Meanwhile,~\cite{RAQ_MIMO_Multiband} proposed a Rydberg atomic quantum-MIMO (RAQ-MIMO) architecture, demonstrating that a vapor-cell-based array can concurrently process multiple RF bands using quantum transconductance modeling and weighted minimum mean-square error (WMMSE) optimization. Collectively, these results suggest that RARs are rapidly emerging as a compelling candidate for 6G receiver front-ends.


Recent studies have begun to extend the applicability of RARs to emerging 6G scenarios; however, such efforts remain in their infancy. The work in~\cite{RAR_Classical_Comm_Sensing} provided a demonstration of practical feasibility by experimentally showing that a single RAR can simultaneously perform microwave radar sensing and wireless communication within an integrated hardware platform. Building on this foundation,~\cite{New_Paradigm_RAR_ISAC} conceptually introduced a RAR-enabled ISAC paradigm, offering high-level discussions on channel and waveform characteristics as well as potential transceiver architectures. Nevertheless, the study remained largely visionary, without presenting a concrete optimization framework or analytically validated ISAC performance. More recently,~\cite{RIS_atomic_MIMO} examined RIS-RAR integration, demonstrating that constructive field shaping via RIS helps suppress photodetector (PD)-related nonlinear distortion and improves demodulation reliability. Taken together, these works indicate that while feasibility demonstrations, conceptual ISAC visions, and preliminary RIS-assisted enhancements have emerged independently, there is, to the best of our knowledge, no rigorous formulation or system-level study that \textit{unifies RAR and RIS-ISAC within one comprehensive framework.}

Motivated by these challenges, this paper presents a unified RIS-ISAC framework that explicitly incorporates multiple RAR-equipped users. By leveraging the field-shaping and combining properties of RIS, the system can increase the effective field strength and improve spatial field conditioning at each RAR, thereby allowing the intrinsic quantum-level sensitivity of the RAR to be more fully utilized. Such controllability is also particularly beneficial in ISAC, which in turn leads us to develop a new formulation that captures RIS-ISAC under multi-RAR reception and propose a sum-rate-oriented utility maximization problem, which is efficiently solved via a tailored block coordinate descent (BCD) framework. Our main contributions are summarized as follows:
\begin{itemize}
\item We establish a downlink RIS-ISAC model with multiple RAR-aided users, jointly capturing: (i) Rydberg-based optical readout with LO-assisted signal injection for effective symbol detection, and (ii) RIS-coordinated signal construction that controls both reflected and direct-path components to enable ISAC operation. The RAR channel is reformulated into an equivalent multi-cell real-valued sensing-detection model, enabling closed-form SINR characterization for ISAC optimization.
\item To balance dual ISAC objectives, we formulate a CRB-aware weighted sum-utility maximization problem that jointly accounts for radar estimation accuracy and multi-user downlink throughput. The optimization determines the precoding matrix and RIS configuration under transmit-power, unit-modulus RIS, and RAR-induced magnitude-only detectability constraints.
\item We design a scalable BCD-based solution framework that decouples the problem via: (i) fractional programming (FP) for the communication objective, (ii) CRB relaxation for tractable sensing integration, and (iii) majorization-minimization (MM) and Alternating Direction Method of Multipliers (ADMM)-based decomposition for joint precoder and RIS phase optimization.
\item Extensive simulations demonstrate that the proposed framework significantly outperforms benchmark schemes across diverse operating conditions, achieving near communication-only performance while maintaining strong sensing accuracy, thereby validating the effectiveness of multi-RAR reception in RIS-ISAC systems.
\end{itemize}
\emph{The key novelty of this paper lies not in RIS or RAR individually, but in revealing how RIS-enabled field shaping fundamentally changes the CRB-rate tradeoff under magnitude-only quantum receivers.}

\begin{figure}[t]
	\begin{center}
		\includegraphics[width=0.9\columnwidth,keepaspectratio]%
		{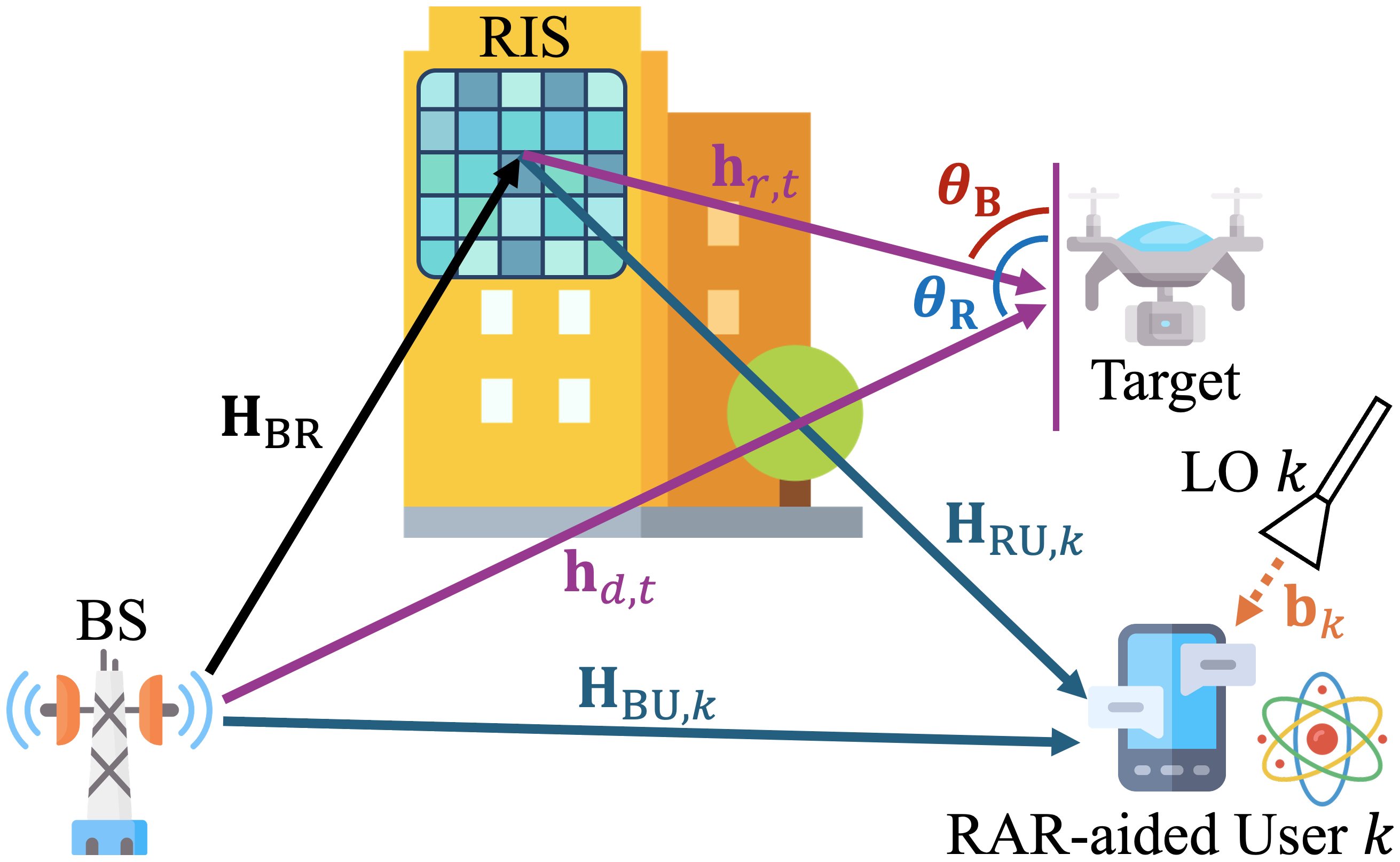}
		\caption{An RIS-ISAC system with $K$ RAR-aided users.}
		\label{fig_sys}
	\end{center}
\end{figure}
\section{System Model}
\label{sec:sys_model}
\subsection{Overall ISAC Architecture}
As illustrated in Fig.~\ref{fig_sys}, we consider a narrowband RIS-ISAC system, where a dual-functional BS performs downlink communications to $K$ users and monostatic radar sensing of a remote target. The system consists of
\begin{itemize}
  \item A BS equipped with $N_t$ RF antennas with half-wavelength spacing. 
  Leveraging advanced self-interference (SI) suppression~\cite{kwakfd}, the BS operates in full-duplex mode with ideal cancellation of SI.
    \item An $N$-element RIS with phase shifts described by $\boldsymbol\Phi$.
  \item $K$ downlink users, where each user employs an $M$-cell RAR and an LO, and a single point target.
\end{itemize}
The RIS is modeled by the reflection matrix $\boldsymbol\Phi\triangleq\mathrm{diag}\big(\{e^{j\theta_n}\}_{n=1}^N\big)~(\forall\theta_n\in[0,2\pi))$, and let $\mathbf s_c \triangleq [s_{c,1} \cdots s_{c,K}]^{\mathrm T}\in\mathbb C^{K}$ the symbol vector with $s_{c,k}$ to user $k$ with $ \mathbb E[\mathbf s_c\mathbf s_c^*]  = \mathbf I_K$, where $\mathbf{I}_{(\cdot)}$ is an identity matrix with corresponding size of row. The BS applies a precoder $\mathbf W\in\mathbb C^{N_t\times K}$ and transmits $\mathbf x$ that satisfies:
\begin{equation}
  \mathbf x
  \triangleq
  \mathbf W\mathbf s_c
  \in\mathbb C^{N_t},  ~  \mathbf R_x \triangleq \mathbb E[\mathbf x\mathbf x^*]  = \mathbf W\mathbf W^*.
  \label{eq:tx_signal}
\end{equation}
The BS transmit power is hence limited by the budget $P_{\max}$:
\begin{equation}
  \mathrm{tr}(\mathbf R_x) =\mathrm{tr}(\mathbf W\mathbf W^*) \le P_{\max}.
  \label{eq:power_constraint}
\end{equation}
We let $\mathbf H_{\mathrm{BR}}\in\mathbb C^{N\times N_t}$ the BS-RIS channel and $\mathbf H_{\mathrm{RU},k}\in\mathbb C^{M\times N}$ and $\mathbf H_{\mathrm{BU},k}\in\mathbb C^{M\times N_t}$ the RIS- and direct BS-RAR channel of user $k$, respectively. The elements of the atomic channels follow a multipath Rydberg EM coupling model~\cite{Precoding_atomicMIMO}, which characterizes the physical interaction between the incident multipath EM fields and the Rydberg atomic transitions. In particular, the $(m,n)$th entry of $\mathbf H_{\mathrm{RU},k}$ is given by~\cite{atomicjsac}
\begin{equation}
  [\mathbf H_{\mathrm{RU},k}]_{m,n}  =  \sum_{\ell=1}^{L_0}  \frac{1}{\hbar}  \boldsymbol\mu_{\mathrm{eg}}^{\mathrm T}\boldsymbol\epsilon_{m,n,k,\ell} \rho_{m,n,k,\ell}  e^{j\phi_{m,n,k,\ell}},
  \label{eq:H_RU_elem}
\end{equation}
where $L_0$ is the number of multipaths, $\hbar$ is the reduced Planck constant, $\boldsymbol\mu_{\mathrm{eg}}\in\mathbb R^{3}$ is the electric dipole moment of the Rydberg RF transition, and $\boldsymbol\epsilon_{m,n,k,\ell}\in\mathbb R^{3}$, $\rho_{m,n,k,\ell}$, $\phi_{m,n,k,\ell}$ denote, respectively, the polarization, path attenuation, and phase of the $\ell$th multipath component between RIS element $n$ and vapor cell $m$ of user $k$. The same structure applies to $\{\mathbf H_{\mathrm{BU},k}\}$ with appropriate indices and propagation parameters.

For a given $\boldsymbol\Phi$, the effective BS-RAR channel for user $k$ is
\begin{equation}
  \mathbf H_{\mathrm{eff},k}^{\mathrm{com}}(\boldsymbol\Phi)
  \triangleq
  \mathbf H_{\mathrm{RU},k}\boldsymbol\Phi\mathbf H_{\mathrm{BR}}  + \mathbf H_{\mathrm{BU},k}  \in\mathbb C^{M\times N_t}.
  \label{eq:Heff_com_k}
\end{equation}
Therefore, the complex RF field $\mathbf r_k^{\mathrm{com}}$ at the input of user $k$ is
\begin{equation}
  \mathbf r_k^{\mathrm{com}}  =   \mathbf H_{\mathrm{eff},k}^{\mathrm{com}}(\boldsymbol\Phi)\mathbf x   + \mathbf b_k + \mathbf n_k,
  \label{eq:r_k_com}
\end{equation}
where $\mathbf b_k\in\mathbb C^{M}$ denotes the RF field of LO corresponds to user $k$, and $\mathbf n_k\sim\mathcal{CN}(\mathbf 0,\sigma_{q,k}^2\mathbf I_M)$ is the measurement noise. The $m$th LO component $b_{k,m}$ at user $k$ is modeled as
\begin{equation}
  b_{k,m}  =  \frac{s_b}{\hbar}  \boldsymbol\mu_{\mathrm{eg}}^{\mathrm T}  \boldsymbol\epsilon_{k,m,b}  \rho_{k,m,b}  \sqrt{P_b}e^{j\phi_{k,m,b}}  = |b_{k,m}|e^{j\angle b_{k,m}},
  \label{eq:b_k_m}
\end{equation}
where $s_b$ is a known reference symbol, $\angle b_{k,m}$ is the phase of $b_{k,m}$, $P_b$ is the LO power and $\boldsymbol\epsilon_{k,m,b}$, $\rho_{k,m,b}$, $\phi_{k,m,b}$ represent its polarization direction, path loss, and phase, respectively~\cite{Precoding_atomicMIMO,RIS_atomic_MIMO,atomicjsac}.
\begin{figure}[t]
	\begin{center}
		\includegraphics[width=0.99\columnwidth,keepaspectratio]%
		{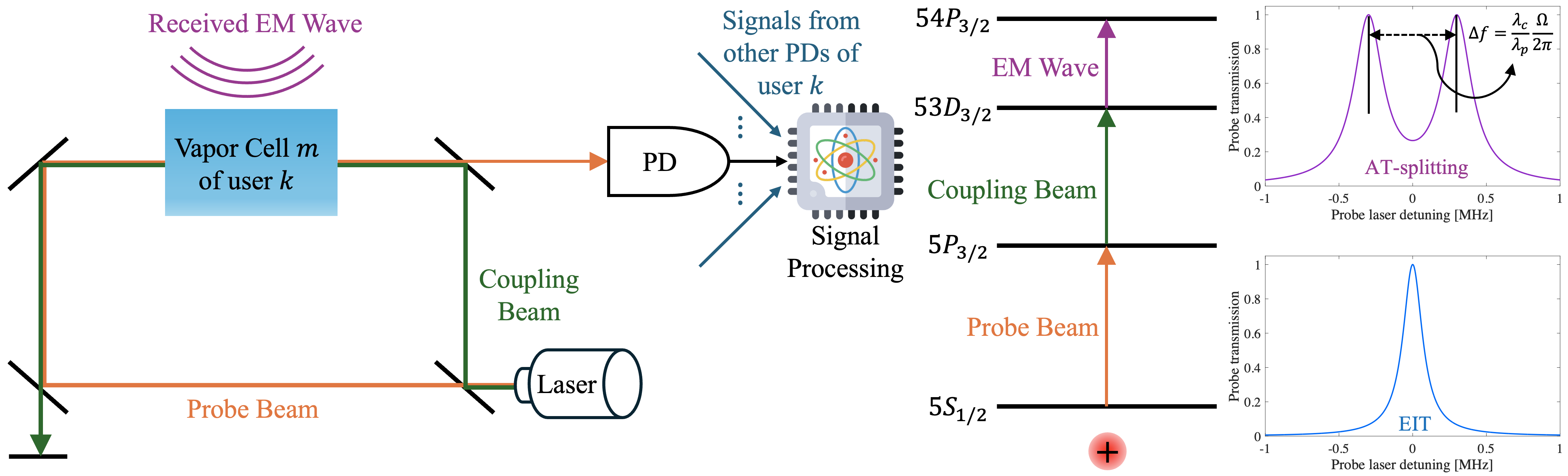}
		\caption{Illustration of signal processing in RAR. Specifically, the incident EM wave couples two highly excited Rydberg states (e.g., $53D_{3/2}$ and $54P_{3/2}$), giving rise to the AT-splitting phenomenon. The resulting spectral separation $\Delta f$ is subsequently mapped to the corresponding Rabi frequency $\Omega$ via~\eqref{rabidef}.}
		\label{fig_rabi}
	\end{center}
\end{figure}
\subsection{Optical Readout at $K$ RAR-Aided Users}
As depicted in Fig.~\ref{fig_rabi}, the incident RF field couples with the Rydberg atomic transitions, giving rise to EIT and AT-splitting~\cite{efm}. The AT-splitting interval $\{\Delta f_m\}$ directly maps to the corresponding Rabi frequency $\{\Omega_m\}$~\cite{Precoding_atomicMIMO, fox}:
\begin{equation}
\label{rabidef}
\Delta f_m=\frac{\lambda_c}{\lambda_p}\frac{\Omega_m}{2\pi}~(\forall m),
\end{equation}
where $\lambda_c$ and $\lambda_p$ denote the wavelengths of the coupling and probe beams, respectively. Hence, by sensing $\Delta f_m$, one can retrieve $\Omega_m$ by~\eqref{rabidef}, which subsequently enables the recovery of the transmitted symbol since $\{\Omega_m\}$ equals to the measurement of $\mathbf r_k^{\mathrm{com}}$~\cite{qmusic}. In particular, the measurement quantity is expressed as~\cite{RIS_atomic_MIMO,atomicjsac}
\begin{equation}
\tilde{\mathbf z}_k  \triangleq  |\mathbf r_k^{\mathrm{com}}|  =  \big|\mathbf H_{\mathrm{eff},k}^{\mathrm{com}}(\boldsymbol\Phi)\mathbf x + \mathbf b_k + \mathbf n_k  \big|,
\label{eq:z_tilde_k_mag}
\end{equation}
where the magnitude is taken elementwise, and each real-valued entry corresponds to $\{\Omega_m\}$.

To obtain a tractable baseband model, we invoke the strong-reference approximation: $|b_{k,m}|  \gg  \big| [\mathbf H_{\mathrm{eff},k}^{\mathrm{com}}(\boldsymbol\Phi)\mathbf x]_m+ n_{k,m}  \big|~ (\forall m)$, where the LO is positioned in close proximity to the RAR (e.g., typically at distances on the order of centimeters~\cite{rarclose3, Harnessing_RAR_Tutorial, Precoding_atomicMIMO}), making this separation much shorter than the BS-RAR distance. Thereafter subtracting $|b_{k,m}|$, the photocurrent $z_{k,m}$ at cell $m$ of user $k$ is approximated as a real-part detection model~\cite{Precoding_atomicMIMO}:
\begin{equation}
\begin{aligned}
  z_{k,m}  &\triangleq   \tilde z_{k,m} - |b_{k,m}|\approx \Re\Big\{ \big[ \mathbf H_{\mathrm{eff},k}^{\mathrm{com}}(\boldsymbol\Phi)\mathbf x \big]_m     e^{-j\angle b_{k,m}} \Big\} + \eta_{k,m},
  \label{eq:z_k_m}
\end{aligned}
\end{equation}
where $\eta_{k,m}\triangleq\Re\{n_{k,m}e^{-j\angle b_{k,m}}\}\sim  \mathcal N\left(0,\frac{\sigma_{q,k}^2}{2}\right)$ is the effective in-phase noise. Stacking all $M$ cells yields
\begin{equation}
  \mathbf z_k  =  \Re\Big\{    \mathbf D_{b,k}    \mathbf H_{\mathrm{eff},k}^{\mathrm{com}}(\boldsymbol\Phi)\mathbf x  \Big\}  +\boldsymbol\eta_k,
  \label{eq:z_k_stack}
\end{equation}
where $\mathbf D_{b,k} \triangleq \mathrm{diag}\big(    e^{-j\angle b_{k,1}},\cdots,e^{-j\angle b_{k,M}}  \big)  \in\mathbb C^{M\times M}$ and $\boldsymbol\eta_k \triangleq [\eta_{k,1}\cdots\eta_{k,M}]^{\mathrm T}  \sim  \mathcal N\left(\mathbf 0,\frac{\sigma_{q,k}^2}{2}\mathbf I_M\right)$. For user $k$, define the LO-phase-aligned effective field matrix $  \widetilde{\mathbf E}_k^{\mathrm{com}}(\boldsymbol\Phi)
  \triangleq
  \mathbf D_{b,k}
  \mathbf H_{\mathrm{eff},k}^{\mathrm{com}}(\boldsymbol\Phi)
  \in\mathbb C^{M\times N_t}$. Substituting it into~\eqref{eq:z_k_stack} yields
\begin{equation}
  \mathbf z_k
  =
  \Re\big\{
    \widetilde{\mathbf E}_k^{\mathrm{com}}(\boldsymbol\Phi)\mathbf x
  \big\}
  + \boldsymbol\eta_k,
  \label{eq:z_k_u}
\end{equation}
and by stacking all $K$ user RAR outputs, we get
\begin{equation}
\label{krar}
  \mathbf z_{\mathrm{com}}  \triangleq
  \begin{bmatrix}
  \mathbf z_1\\
  \vdots\\
  \mathbf z_K
  \end{bmatrix} \in\mathbb R^{KM},  ~
  \boldsymbol\eta_{\mathrm{com}}\triangleq
  \begin{bmatrix}
  \boldsymbol\eta_1\\
  \vdots\\
  \boldsymbol\eta_K
  \end{bmatrix} \in\mathbb R^{KM},
\end{equation}
and
\begin{equation}
  \widetilde{\mathbf E}_{\mathrm{com}}(\boldsymbol\Phi)  \triangleq
  \begin{bmatrix}
  \widetilde{\mathbf E}_1^{\mathrm{com}}(\boldsymbol\Phi)\\
  \vdots\\
  \widetilde{\mathbf E}_K^{\mathrm{com}}(\boldsymbol\Phi)
  \end{bmatrix}  \in\mathbb C^{KM\times N_t},
  \label{eq:Etilde_com_stack}
\end{equation}
with the global communication-side RAR model
\begin{equation}
  \mathbf z_{\mathrm{com}}
  =
  \Re\big\{
    \widetilde{\mathbf E}_{\mathrm{com}}(\boldsymbol\Phi)\mathbf x
  \big\}
  + \boldsymbol\eta_{\mathrm{com}}.
  \label{eq:z_com_global}
\end{equation}
\subsection{RIS-Aided Monostatic Sensing at BS}
We now model the radar sensing function at the BS, following the RIS-enabled sensing and RIS-ISAC models in~\cite{Song_CRB_IRS,Liu_RIS_ISAC_CRB}. The BS acts as a monostatic MIMO radar: the same $N_t$ antennas are used for transmission and reception. Let $\mathbf h_{r,t}\in\mathbb C^{N}$ and $\mathbf h_{d,t}\in\mathbb C^{N_t}$ the RIS- and the direct BS-target channel, respectively. The target is modeled as a point target with direction-of-arrival (DoA) vector $  \boldsymbol\theta \triangleq
  \begin{bmatrix}
    \theta_{\mathrm{B}} \\ \theta_{\mathrm{R}}
  \end{bmatrix}
  \in\mathbb R^2$, where $\theta_{\mathrm{B}}$ and $\theta_{\mathrm{R}}$ are the DoA of the target with respect to the BS and RIS, respectively. Following~\cite{Song_CRB_IRS, Liu_RIS_ISAC_CRB}, the equivalent BS-RIS-target-RIS-BS channel is
\begin{equation}
  \mathbf H_t(\boldsymbol\Phi,\boldsymbol\theta)
  \triangleq
  \big(
    \mathbf h_{d,t} + \mathbf H_{\mathrm{BR}}^{\mathrm{T}}\boldsymbol\Phi\mathbf h_{r,t}
  \big)
  \big(
    \mathbf h_{d,t}^{\mathrm T} + \mathbf h_{r,t}^{\mathrm T}\boldsymbol\Phi\mathbf H_{\mathrm{BR}}
  \big)
  \in\mathbb C^{N_t\times N_t}.
  \label{eq:H_t_Phi}
\end{equation}
Over a radar dwell time with $L$ snapshots, let $\mathbf X \triangleq [\mathbf x[1]\cdots\mathbf x[L]]\in\mathbb C^{N_t\times L}$ denote the transmitted signal matrix, with sample covariance $ \mathbf R_x
  =
  \frac{1}{L}\sum_{\ell=1}^L \mathbf x[\ell]\mathbf x[\ell]^*
  =
  \frac{1}{L}\mathbf X\mathbf X^*$, consistent with~\eqref{eq:tx_signal}. The received echo $\mathbf Y_r$ at BS is~\cite{Liu_RIS_ISAC_CRB}
\begin{equation}
  \mathbf Y_r=  \alpha_t  \mathbf H_t(\boldsymbol\Phi,\boldsymbol\theta)\mathbf X  + \mathbf N_r,
  \label{eq:Yr_echo}
\end{equation}
where $\alpha_t\in\mathbb C$ is the complex target reflection coefficient and $\mathbf N_r\in\mathbb C^{N_t\times L}$ is the thermal noise matrix with i.i.d. entries $\sim\mathcal{CN}(0,\sigma_r^2)$. Vectorizing~\eqref{eq:Yr_echo}, we obtain
\begin{equation}
  \mathbf y_r
  \triangleq
  \mathrm{vec}(\mathbf Y_r)
  =
  \alpha_t  \mathrm{vec}\big(\mathbf H_t(\boldsymbol\Phi,\boldsymbol\theta)\mathbf X\big)
  + \mathbf n_r,
  \label{eq:y_r_vec}
\end{equation}
where $\mathbf n_r\triangleq \mathrm{vec}(\mathbf N_r)\sim\mathcal{CN}(\mathbf 0,\sigma_r^2\mathbf I_{N_tL})$.
\section{ISAC Problem Formulation}
\label{sec:prob_form}
\subsection{Communication Utility for $K$ RAR-aided Users}
The RAR users detect their information symbols from the in-phase components of $\mathbf z_{\mathrm{com}}$ in~\eqref{eq:z_com_global}. We now develop a communication-theoretic utility based on the SINR achieved at each RAR. Define the LO-phase-aligned effective multiuser channel for user $k$ as
\begin{equation}
  \mathbf G_k(\boldsymbol\Phi,\mathbf W)
  \triangleq
  \mathbf D_{b,k}
  \mathbf H_{\mathrm{eff},k}^{\mathrm{com}}(\boldsymbol\Phi)\mathbf W
  \in\mathbb C^{M\times K},
  \label{eq:Gk_def}
\end{equation}
and let $\mathbf g_{k,i}\in\mathbb C^{M}$ denote its $i$th column, i.e., $\mathbf G_k=[\mathbf g_{k,1}\cdots\mathbf g_{k,K}]$. Then~\eqref{eq:z_k_stack} can be expanded as
\begin{equation}
  \mathbf z_k
  =
  \Re\left\{
    \mathbf g_{k,k} s_{c,k}
    +
    \sum_{i\ne k}\mathbf g_{k,i}s_{c,i}
  \right\}
  + \boldsymbol\eta_k.
  \label{eq:zk_des_int}
\end{equation}
Herein~\eqref{eq:zk_des_int}, the first term corresponds to the desired signal for user $k$, while the summation term represents multiuser interference. The RAR operates cellwise and produces $\mathbf z_k$; we therefore measure signal and interference powers by the
Euclidean norms across the $M$ cells.

Writing $\mathbf g_{k,i}=\mathbf a_{k,i}+j\mathbf b_{k,i}$ with $\mathbf a_{k,i}=\Re\{\mathbf g_{k,i}\}$ and $\mathbf b_{k,i}=\Im\{\mathbf g_{k,i}\}$, and decomposing $s_{c,i}=s_{c,i}^{\mathrm R}+js_{c,i}^{\mathrm I}$ with independent $s_{c,i}^{\mathrm R},s_{c,i}^{\mathrm I}\sim\mathcal N\left(0,\frac{1}{2}\right)$, the in-phase operation in~\eqref{eq:zk_des_int} yields $  \Re\{\mathbf g_{k,i}s_{c,i}\}  =  \mathbf a_{k,i}s_{c,i}^{\mathrm R}  - \mathbf b_{k,i}s_{c,i}^{\mathrm I}$, so that both the real and imaginary parts of $\mathbf g_{k,i}$ contribute to the in-phase photocurrent through the in-phase and quadrature data components. Taking expectations with respect to the symbols, the average desired signal power $P_{k}^{\mathrm{sig}}$ at user $k$ is
\begin{equation}
\begin{aligned}
  P_{k}^{\mathrm{sig}}
  &\triangleq
  \mathbb E\big[
    \|\Re\{\mathbf g_{k,k}s_{c,k}\}\|_2^2
  \big]
  =
  \frac{1}{2}\|\mathbf g_{k,k}\|_2^2.
  \label{eq:Pk_sig}
  \end{aligned}
\end{equation}
Similarly, the average multiuser interference power $P_{k}^{\mathrm{int}}$ is
\begin{equation}
  P_{k}^{\mathrm{int}}
  \triangleq
  \sum_{i\ne k}
  \mathbb E\big[
    \|\Re\{\mathbf g_{k,i}s_{c,i}\}\|_2^2
  \big]
  =
  \frac{1}{2}\sum_{i\ne k}\|\mathbf g_{k,i}\|_2^2,
  \label{eq:Pk_int}
\end{equation}
while the effective noise power is
\begin{equation}
  P_{k}^{\mathrm{noise}}
  \triangleq
  \mathbb E\big[\|\boldsymbol\eta_k\|_2^2\big] =  \frac{M\sigma_{q,k}^2}{2}.
  \label{eq:Pk_noise}
\end{equation}
Collecting~\eqref{eq:Pk_sig}-\eqref{eq:Pk_noise}, we define the SINR of user $k$ as
\begin{equation}
  \mathrm{SINR}_k(\mathbf W,\boldsymbol\Phi)
  \triangleq
  \frac{
    \|\mathbf g_{k,k}(\boldsymbol\Phi,\mathbf W)\|_2^2
  }{\sum_{i\ne k}\|\mathbf g_{k,i}(\boldsymbol\Phi,\mathbf W)\|_2^2
    +
    {M\sigma_{q,k}^2}
  },
  \label{eq:SINRk_def}
\end{equation}
and the overall communication utility $\mathcal U_{\mathrm{com}}$ is expressed as a form of sum-rate in nats/s/Hz:
\begin{equation}
  \mathcal U_{\mathrm{com}}(\mathbf W,\boldsymbol\Phi)\triangleq  \sum_{k=1}^{K}  \ln\big(1+\mathrm{SINR}_k(\mathbf W,\boldsymbol\Phi)\big),
  \label{eq:Ucom}
\end{equation}
hence maximizing $\mathcal U_{\mathrm{com}}$ enhances the communication performance of the multiple-RARs while explicitly accounting for multiuser interference and the RAR noise characteristics.
\begin{remark}
\label{r1r1}
Following~\cite{Precoding_atomicMIMO}, the RAR channel can be transformed into an equivalent real-valued MIMO representation. However, it is derived under a single-user setting with  waterfilling-based input covariance assumptions, and thus cannot be directly applied to the considered multi-RAR scenario. (Indeed, the information-theoretic capacity characterization of multi-RAR systems remains an open research problem.) In contrast, the proposed $\mathcal{U}_{\mathrm{com}}$ is developed for a multi-RAR configuration and is constructed based on an SINR-driven achievable-rate expression, and can be interpreted as an achievable rate under mismatched decoding, which serves as a tractable lower-bound-style performance indicator. From this viewpoint, the use $\mathcal{U}_{\mathrm{com}}$ is practically well-motivated: it captures the throughput performance under realistic RAR architectures experiencing multi-cell interference, enables optimization-friendly analytical expressions, and reflects implementable communication performance.
\end{remark}
\subsection{Cram\'er-Rao Bound of Sensing Performance}
We focus on a point target whose location is parameterized by $\boldsymbol\theta$. We adopt standard line-of-sight (LoS) models for BS/RIS-target links~\cite{Song_CRB_IRS,Liu_RIS_ISAC_CRB}: $\mathbf h_{d,t}(\theta_{\mathrm{B}})  = \beta_d\mathbf a_{\mathrm{B}}(\theta_{\mathrm{B}})     \in\mathbb C^{N_t}, \mathbf h_{r,t}(\theta_{\mathrm{R}})  = \beta_r\mathbf a_{\mathrm{R}}(\theta_{\mathrm{R}})     \in\mathbb C^{N}$, where $\beta_d.~\beta_r\in\mathbb C$ capture path loss and $\mathbf a_{\mathrm{B}}(\theta_{\mathrm{B}})\in\mathbb C^{N_t}$ and $\mathbf a_{\mathrm{R}}(\theta_{\mathrm{R}})\in\mathbb C^{N}$ denote the BS and RIS steering vectors, respectively. We assume $\mathbf H_{\mathrm{BR}}\in\mathbb C^{N\times N_t}$ follows Rician fading:
\begin{equation}
  \mathbf H_{\mathrm{BR}}
  =
  \beta_t\sqrt{\frac{\kappa}{\kappa+1}}
  \mathbf H_{\mathrm{BR},\mathrm{LoS}}
  +
  \beta_t\sqrt{\frac{1}{\kappa+1}}
  \mathbf H_{\mathrm{BR},\mathrm{NLoS}}
  \label{eq:Gt_Rician}
\end{equation}
where $\beta_t$ is the path loss and $\kappa\ge0$ is the Rician $K$-factor. Herein, the LoS component $ \mathbf H_{\mathrm{BR},\mathrm{LoS}}$ is modeled geometrically as $  \mathbf H_{\mathrm{BR},\mathrm{LoS}}  =
  \mathbf a_{\mathrm{R}}(\varphi_{\mathrm{R},t})
  \mathbf a_{\mathrm{B}}^*(\varphi_{\mathrm{B},t})$, 
where $\varphi_{\mathrm{B},t}$ and $\varphi_{\mathrm{R},t}$ are the BS-RIS direction-of-departure (DoD) and DoA, respectively. Thereafter, $\mathbf H_{\mathrm{BR},\mathrm{NLoS}}$ is modeled as the Rayleigh fading components with $\mathcal{CN}(0,1)$ each. Therefore, $\mathbf H_t$ in~\eqref{eq:H_t_Phi} becomes a parametric function of $(\boldsymbol\Phi,\boldsymbol\theta)$:
\begin{equation}
\begin{aligned}
\mathbf H_t(\boldsymbol\Phi,\boldsymbol\theta)= \mathbf v_t(\boldsymbol\Phi,\boldsymbol\theta)
     \mathbf v_t^{\mathrm T}(\boldsymbol\Phi,\boldsymbol\theta)
     \in\mathbb C^{N_t\times N_t},
  \label{eq:Ht_param}
\end{aligned}
\end{equation}
where $\mathbf v_t(\boldsymbol\Phi,\boldsymbol\theta)
  \triangleq
  \mathbf h_{d,t}(\theta_{\mathrm{B}})
  + \mathbf H_{\mathrm{BR}}^{\mathrm{T}}\boldsymbol\Phi\mathbf h_{r,t}(\theta_{\mathrm{R}})$. The derivatives of $\mathbf H_t(\boldsymbol\Phi,\boldsymbol\theta)$ with respect to $\boldsymbol\theta$ will be needed for the Fisher Information Matrix (FIM):
\begin{equation}
  \frac{\partial\mathbf H_t}{\partial\theta_{\mathrm{B}}}
  = \frac{\partial\mathbf v_t}{\partial\theta_{\mathrm{B}}}
     \mathbf v_t^{\mathrm T}
     + \mathbf v_t
       \Big(
         \frac{\partial\mathbf v_t}{\partial\theta_{\mathrm{B}}}
       \Big)^{\mathrm T},
  \frac{\partial\mathbf H_t}{\partial\theta_{\mathrm{R}}}
  = \frac{\partial\mathbf v_t}{\partial\theta_{\mathrm{R}}}
     \mathbf v_t^{\mathrm T}
     + \mathbf v_t
       \Big(
         \frac{\partial\mathbf v_t}{\partial\theta_{\mathrm{R}}}
       \Big)^{\mathrm T},
  \label{eq:Hdot_R}
\end{equation}
with $  \frac{\partial\mathbf v_t}{\partial\theta_{\mathrm{B}}}
  = \beta_d
     \frac{\partial\mathbf a_{\mathrm{B}}(\theta_{\mathrm{B}})}
          {\partial\theta_{\mathrm{B}}}$ and $\frac{\partial\mathbf v_t}{\partial\theta_{\mathrm{R}}}
= \mathbf H_{\mathrm{BR}}^{\mathrm{T}}\boldsymbol\Phi
     \beta_r
     \frac{\partial\mathbf a_{\mathrm{R}}(\theta_{\mathrm{R}})}
          {\partial\theta_{\mathrm{R}}}$.

Recall~\eqref{eq:Yr_echo} and its vectorization in~\eqref{eq:y_r_vec}. Equivalently, for CRB derivation it is convenient to isolate the dependence on $\boldsymbol\theta$ and $\alpha_t$ via a deterministic steering vector~\cite{Liu_RIS_ISAC_CRB}. Let $
  \mathbf h(\boldsymbol\Phi,\boldsymbol\theta)
  \triangleq
  \mathrm{vec}\big(\mathbf H_t(\boldsymbol\Phi,\boldsymbol\theta)\mathbf X\big)
  \in\mathbb C^{N_tL}$, so that
\begin{equation}
  \mathbf y_r
  =
  \boldsymbol\mu(\boldsymbol\xi)
  + \mathbf n_r,
  ~
  \boldsymbol\mu(\boldsymbol\xi)
  \triangleq
  \alpha_t \mathbf h(\boldsymbol\Phi,\boldsymbol\theta),
  \label{eq:y_mean_def}
\end{equation}
and $\boldsymbol\xi$ is the real parameter vector to be estimated:
\begin{equation}
  \boldsymbol\xi
  \triangleq
  \begin{bmatrix}
    \boldsymbol\theta \\
    \alpha_{\mathrm{R}} \\
    \alpha_{\mathrm{I}}
  \end{bmatrix}
  \in\mathbb R^{4},
\end{equation}
where $\alpha_{\mathrm R}\triangleq\Re\{\alpha_t\}$ and $\alpha_{\mathrm I}\triangleq\Im\{\alpha_t\}$. Under~\eqref{eq:y_mean_def}, the FIM associated with $\boldsymbol\xi$ is given by~\cite{Song_CRB_IRS,Liu_RIS_ISAC_CRB, Kay}
\begin{equation}
  \mathbf J(\mathbf W,\boldsymbol\Phi,\boldsymbol\xi)
  =
  \frac{2}{\sigma_r^2}
  \Re\bigg\{
    \Big(
      \frac{\partial\boldsymbol\mu}{\partial\boldsymbol\xi}
    \Big)^*
    \Big(
      \frac{\partial\boldsymbol\mu}{\partial\boldsymbol\xi}
    \Big)
  \bigg\},
  \label{eq:FIM_general}
\end{equation}
where $\frac{\partial\boldsymbol\mu}{\partial\boldsymbol\xi}$ is the Jacobian of $\boldsymbol\mu(\boldsymbol\xi)$. The partial derivatives of $\boldsymbol\mu$ with respect to the components of $\boldsymbol\xi$ are
\begin{equation}
\begin{aligned}
  \frac{\partial\boldsymbol\mu}{\partial\theta_i}
  = \alpha_t
     \frac{\partial\mathbf h(\boldsymbol\Phi,\boldsymbol\theta)}
          {\partial\theta_i},
  \frac{\partial\boldsymbol\mu}{\partial\alpha_{\mathrm{R}}}
  = \mathbf h(\boldsymbol\Phi,\boldsymbol\theta),\frac{\partial\boldsymbol\mu}{\partial\alpha_{\mathrm{I}}}= j\mathbf h(\boldsymbol\Phi,\boldsymbol\theta),
  \label{eq:dmu_dtheta}
\end{aligned}
\end{equation}
where $i\in\{\mathrm{B},\mathrm{R}\}$ and $
  \frac{\partial\mathbf h(\boldsymbol\Phi,\boldsymbol\theta)}
       {\partial\theta_i}
  =
  \mathrm{vec}\Big(
    \frac{\partial\mathbf H_t(\boldsymbol\Phi,\boldsymbol\theta)}
            {\partial\theta_i}
    \mathbf X
  \Big)$. Plugging~\eqref{eq:dmu_dtheta} into~\eqref{eq:FIM_general}, $\mathbf J$ takes the block-partitioned form:
\begin{equation}
  \mathbf J
  =
  \begin{bmatrix}
    \mathbf J_{\boldsymbol\theta\boldsymbol\theta}
    &
    \mathbf J_{\boldsymbol\theta\boldsymbol\alpha}
    \\
    \mathbf J_{\boldsymbol\alpha\boldsymbol\theta}
    &
    \mathbf J_{\boldsymbol\alpha\boldsymbol\alpha}
  \end{bmatrix},
  \label{eq:FIM_block}
\end{equation}
where $\boldsymbol\alpha\triangleq[\alpha_{\mathrm{R}},\alpha_{\mathrm{I}}]^{\mathrm T}$ and each block is given in~\eqref{jta}
\begin{figure*}
\begin{equation}
\begin{aligned}
  [\mathbf J_{\boldsymbol\theta\boldsymbol\theta}]_{i,j}  &=  \frac{2}{\sigma_r^2}  \Re\Big\{    \alpha_t^*    \alpha_t    \bigg(      \frac{\partial\mathbf h}{\partial\theta_i}    \bigg)^*
    \frac{\partial\mathbf h}{\partial\theta_j}
  \Big\},  \mathbf J_{\boldsymbol\alpha\boldsymbol\alpha}
  =
  \frac{2}{\sigma_r^2}
  \Re\Big\{
    \big[\mathbf h~j\mathbf h\big]^*
    \big[\mathbf h~j\mathbf h\big]
  \Big\}=
  \frac{2}{\sigma_r^2}
  \|\mathbf h(\boldsymbol\Phi,\boldsymbol\theta)\|_2^2\mathbf I_2,\\
  \big[\mathbf J_{\boldsymbol\theta\boldsymbol\alpha}\big]_{i,1}
  &=  \frac{2}{\sigma_r^2}  \Re\Big\{    \alpha_t^{*}
    \Big(\frac{\partial\mathbf h(\boldsymbol\Phi,\boldsymbol\theta)}              {\partial\theta_i}\Big)^*
    \mathbf h(\boldsymbol\Phi,\boldsymbol\theta)  \Big\},  \big[\mathbf J_{\boldsymbol\theta\boldsymbol\alpha}\big]_{i,2}=  \frac{2}{\sigma_r^2}  \Re\Big\{    j\alpha_t^{*}    \Big(\frac{\partial\mathbf h(\boldsymbol\Phi,\boldsymbol\theta)}  {\partial\theta_i}\Big)^*    \mathbf h(\boldsymbol\Phi,\boldsymbol\theta)
  \Big\}.
  \label{jta}
\end{aligned}
\end{equation}
\hrule
\end{figure*}
for $i, j\in\{\mathrm B,\mathrm R\}\leftrightarrow\{1, 2\}$, and $\mathbf J_{\boldsymbol\alpha\boldsymbol\theta}=\mathbf J_{\boldsymbol\theta\boldsymbol\alpha}^{\mathrm T}$. 
Hence, the CRB for $\boldsymbol\theta$ is given by the upper-left $2\times2$ block of $\mathbf J^{-1}$~\cite{Kay}:
\begin{equation}
\begin{aligned}
  \mathrm{CRB}_{\boldsymbol\theta}(\mathbf W,\boldsymbol\Phi)
    &=
  \Big(
    \mathbf J_{\boldsymbol\theta\boldsymbol\theta}
    -
    \mathbf J_{\boldsymbol\theta\boldsymbol\alpha}
    \mathbf J_{\boldsymbol\alpha\boldsymbol\alpha}^{-1}
    \mathbf J_{\boldsymbol\alpha\boldsymbol\theta}
  \Big)^{-1},
  \label{eq:CRB_theta_matrix}
  \end{aligned}
\end{equation}
and we consider $\mathrm{tr}(\mathrm{CRB}_{\boldsymbol\theta}(\mathbf W,\boldsymbol\Phi))$ as the overall scalar sensing metric~\cite{Kay}.
\subsection{Joint Precoder and RIS Phase Design Problem}
We finally formulate the problem, where $\mathbf W$ and $\boldsymbol\Phi$ are jointly optimized to strike a trade-off between communication and sensing. To do this, we investigate to maximize $\mathcal U_{\mathrm{com}}$ while satisfying a CRB constraint, the transmit power budget and the unit-modulus constraint of RIS:
\begin{equation}
\label{prob:ISAC_joint}
\begin{aligned}
  \max_{\mathbf W,\boldsymbol\Phi}~
  &  \mathcal U_{\mathrm{com}}(\mathbf W,\boldsymbol\Phi)
  \\
  \mathrm{s.t.}~  & \mathrm{tr}(\mathrm{CRB}_{\boldsymbol\theta}(\mathbf W,\boldsymbol\Phi))<\epsilon,\\
  & \mathrm{tr}(\mathbf W\mathbf W^*) \le P_{\max}, |\boldsymbol\Phi_{n,n}| = 1~ (\forall n\in\{1,\cdots,N\}).
\end{aligned}
\end{equation}
Problem~\eqref{prob:ISAC_joint} is nonconvex due to the fractional SINRs, the inverse FIM structure, and the unit-modulus constraint on $\boldsymbol\Phi$. In the subsequent sections, we develop the joint optimization framework with respect to $\mathbf W$ and $\boldsymbol\Phi$, and devise a tractable BCD framework to handle the associated nonconvexity.
\section{Joint Optimization Framework}
\label{sec:algo}
\subsection{FP-Based Transformation of $\mathcal U_{\mathrm{com}}$}
Following the FP framework in~\cite{fracp} with applying the Lagrangian-dual reformulation, we introduce an auxiliary vector $\mathbf r = [r_1\cdots r_K]^{\mathrm T}\in\mathbb R_{+}^{K}$ to rewrite each logarithmic term:
\begin{equation}
\begin{aligned}
\label{eq:FP_step1}
&\ln(1+\mathrm{SINR}_k)\\
&=\max_{r_k\ge0}\Big(\ln(1+r_k)-r_k+(1+r_k)
\frac{
  \|\mathbf g_{k,k}\|_2^2
}{
  \sum_{j=1}^{K}\|\mathbf g_{k,j}\|_2^2
  + M\sigma_{q,k}^2
}\Big),
\end{aligned}
\end{equation}
where maximum is achieved at
\begin{equation}
\label{rkmax}
r_k^\star=\frac{\|\mathbf g_{k,k}\|_2^2
}{
\sum_{j\neq k}\|\mathbf g_{k,j}\|_2^2
+ M\sigma_{q,k}^2
}=\mathrm{SINR}_k.
\end{equation}
Accordingly, $\mathcal U_{\mathrm{com}}$ becomes maximizing
\begin{equation}
\begin{aligned}
  \mathcal F_1(\mathbf W,\boldsymbol\Phi,\mathbf r)
  =&\sum_{k=1}^{K}
  \ln(1+r_k)
  -
  \sum_{k=1}^{K} r_k
  \\
  &+
  \sum_{k=1}^{K}
  (1+r_k)
  \frac{
    \|\mathbf g_{k,k}\|_2^2
  }{
    \sum_{j=1}^{K}\|\mathbf g_{k,j}\|_2^2
    + M\sigma_{q,k}^2
  }
\end{aligned}
\label{eq:Ucom_FP1}
\end{equation}
over $\mathbf r \ge \mathbf 0$, with maximum at~\eqref{rkmax}. To eliminate the fractional structure, we introduce another set of auxiliary vectors $\{\mathbf c_k\}_{k=1}^K$ with $\mathbf c_k\in\mathbb C^{M}$.
For each user $k$, define
\begin{equation}
  B_k(\mathbf W,\boldsymbol\Phi)
  \triangleq
  \sum_{j=1}^{K}\|\mathbf g_{k,j}(\boldsymbol\Phi,\mathbf W)\|_2^2
  + M\sigma_{q,k}^2.
  \label{eq:Bk_def}
\end{equation}
Then, using the quadratic transform identity~\cite{fracp}
\begin{equation}
  (1+r_k)\frac{\|\mathbf g_{k,k}\|_2^2}{B_k}
  =
  \max_{\mathbf c_k\in\mathbb C^{M}}
  \Big(
    2\sqrt{1+r_k}
    \Re\{\mathbf c_k^{*}\mathbf g_{k,k}\}-B_k\|\mathbf c_k\|_2^2
  \Big),
  \label{qtivec}
\end{equation}
with maximum at
\begin{equation}
\label{ckmax}
\mathbf c_k^\star=\frac{\sqrt{1+r_k}}{B_k}\mathbf g_{k,k},
\end{equation}
we obtain the following equivalent FP reformulation of $\mathcal U_{\mathrm{com}}$ by maximizing
\begin{equation}
\begin{aligned}
&\mathcal F(\mathbf W,\boldsymbol\Phi,\mathbf r,\{\mathbf c_k\})\triangleq\sum_{k=1}^{K}
\ln(1+r_k)
-
\sum_{k=1}^{K}r_k+\\
&\sum_{k=1}^{K}
\bigg(
  2\sqrt{1+r_k}
  \Re\big\{\mathbf c_k^{*}\mathbf g_{k,k}\big\}
 -\|\mathbf c_k\|_2^2
  \Big(
    \sum_{j=1}^{K}\|\mathbf g_{k,j}\|_2^2
    + M\sigma_{q,k}^2
  \Big)
\bigg).
\end{aligned}
\label{eq:Ucom_FP2_vector}
\end{equation}
over $\mathbf r \ge \mathbf 0$ and $\mathbf c_k \in \mathbb C^{M}$. Let $\mathbf w \triangleq \mathrm{vec}(\mathbf W)\in\mathbb C^{N_tK}$ denote the stacked precoder, and collect the auxiliary vectors as $\mathbf C\triangleq[\mathbf c_1 \cdots \mathbf c_K]\in\mathbb C^{M\times K}$. Then all $\mathbf W$-dependent terms in~\eqref{eq:Ucom_FP2_vector} can be written as a quadratic function of $\mathbf w$. Specifically, there exist a vector $\mathbf a$, a matrix $\mathbf B$, and a scalar $\epsilon_1$ such that:
\begin{equation}
  \mathcal F(\mathbf W,\boldsymbol\Phi,\mathbf r,\{\mathbf c_k\})
  =
  \Re\{\mathbf a^{*}\mathbf w\}
  -
  \|\mathbf B\mathbf w\|_2^2
  +
  \epsilon_1,
  \label{eq:FP_compact_w_vector_final}
\end{equation}
which is conditionally concave in $\mathbf w$ for fixed $(\boldsymbol\Phi,\mathbf r,\{\mathbf c_k\})$, where $\mathbf a$ and $\mathbf B$ is given by
\begin{equation}
\label{defab}
\mathbf a
=
2\sum_{k=1}^K
\sqrt{1+r_k}\mathbf J_{k,k}^{*}\mathbf c_k, \mathbf B
\triangleq
\begin{bmatrix}
\|\mathbf c_1 \|_2\mathbf{\widetilde J}_1\\
\vdots\\
\|\mathbf c_K \|_2\mathbf{\widetilde J}_K
\end{bmatrix}
\end{equation}
with
\begin{equation}
\label{defdef1}
\begin{aligned}
&\mathbf J_{k,i}
\triangleq
\widetilde{\mathbf E}_k^{\mathrm{com}}(\boldsymbol\Phi)\mathbf S_i
\in\mathbb C^{M\times N_tK},\\
&\mathbf S_i
\triangleq
\big[
\mathbf 0_{N_t\times N_t(i-1)}~
\mathbf I_{N_t}~
\mathbf 0_{N_t\times N_t(K-i)}
\big]\in\mathbb R^{N_t \times N_t K},\\
&
\mathbf{\widetilde J}_k
\triangleq
\begin{bmatrix}
\mathbf J_{k,1}\\
\vdots\\
\mathbf J_{k,K}
\end{bmatrix}
 \in\mathbb C^{KM\times N_tK}.
\end{aligned}
\end{equation}
Likewise, for fixed $\mathbf W$, $\mathbf r$, and $\{\mathbf c_k\}$, the terms depending on the RIS phase vector $\boldsymbol\phi \triangleq [e^{j\theta_1} \cdots e^{j\theta_N}]^{\mathrm T}~(\boldsymbol\Phi=\mathrm{diag}(\boldsymbol\phi))$ admit a similar quadratic form. That is, there exist $\mathbf g$, a Hermitian matrix $\mathbf D\succeq\mathbf 0$, and a constant $\epsilon_2$ such that
\begin{equation}
  \mathcal F(\mathbf W,\boldsymbol\Phi,\mathbf r,\{\mathbf c_k\})
  =
  \Re\{\mathbf g^{*}\boldsymbol\phi\}
  -
  \boldsymbol\phi^{*}\mathbf D \boldsymbol\phi
  +
  \epsilon_2,
  \label{eq:FP_compact_phi_vector_final}
\end{equation}
which is conditionally concave in $\boldsymbol\phi$ given $(\mathbf W,\mathbf r,\{\mathbf c_k\})$, where $\mathbf g$ and $\mathbf D$ are given by
\begin{equation}
\label{defgD}
\begin{aligned}
&\mathbf g
=
2\Bigg(
  \sum_{k=1}^K
    \sqrt{1+r_k}\mathbf F_{k,k}^{*}\mathbf c_k
  -
  \sum_{k=1}^K
    \|\mathbf c_k \|_2
    \sum_{j=1}^K
      \mathbf F_{k,j}^{*}\mathbf d_{k,j}
\Bigg),\\
&\mathbf D
=
\sum_{k=1}^K
\|\mathbf c_k \|_2
\sum_{j=1}^K
\mathbf F_{k,j}^{*}\mathbf F_{k,j}.
\end{aligned}
\end{equation}
with $\mathbf F_{k,i}
\triangleq
\mathbf D_{b,k}\mathbf H_{\mathrm{RU},k}\mathrm{diag}(\mathbf q_{k,i}), \mathbf q_{k,i}
\triangleq
\mathbf H_{\mathrm{BR}}\mathbf w_i, \mathbf d_{k,i}
\triangleq
\mathbf D_{b,k}\mathbf H_{\mathrm{BU},k}\mathbf w_i$. Therefore, these FP reformulation yields a concave objective with respect to each individual $\mathbf W$, $\boldsymbol\Phi$, $\mathbf r$, and $\{\mathbf c_k\}$ when the others are fixed. This blockwise concavity enables an AO framework given later in Section~\ref{ubub}, where $(\mathbf r,\{\mathbf c_k\})$ admit closed-form updates, while $\mathbf W$ and $\boldsymbol\Phi$ are updated via convex quadratic subproblems.
\subsection{Reformulation of CRB-Constraint}
\label{subsec:W_update_CRB_fixed_Phi}
Thereafter, we reformulate the CRB constraint into a set of linear matrix inequalities (LMIs) that depend on $\mathbf R_x=\mathbf W\mathbf W^*$. We first rewrite $\mathrm{tr}(\mathrm{CRB}_{\boldsymbol\theta}(\mathbf W,\boldsymbol\Phi))<\epsilon$ in~\eqref{prob:ISAC_joint} via an auxiliary positive semidefinite matrix $\mathbf J_{\mathrm {aux}}\in\mathbb R^{2\times2}$. Define
\begin{equation}
  \mathbf A(\mathbf W,\boldsymbol\Phi)
  \triangleq
  \mathbf J_{\boldsymbol\theta\boldsymbol\theta}
  -
  \mathbf J_{\boldsymbol\theta\boldsymbol\alpha}
  \mathbf J_{\boldsymbol\alpha\boldsymbol\alpha}^{-1}
  \mathbf J_{\boldsymbol\alpha\boldsymbol\theta}
  \in\mathbb R^{2\times2},
  \label{eq:A_def_CRB}
\end{equation}
so that $\mathrm{CRB}_{\boldsymbol\theta}(\mathbf W,\boldsymbol\Phi)=\mathbf A(\mathbf W,\boldsymbol\Phi)^{-1}$. Since $\mathbf A(\mathbf W,\boldsymbol\Phi)\succ\mathbf0$ for identifiable parameter vectors, we have $  \mathrm{tr}\big(    \mathbf A(\mathbf W,\boldsymbol\Phi)^{-1}  \big)  \le \epsilon$ iff there exists a positive definite matrix $\mathbf J_{\mathrm {aux}}\succ\mathbf0$ such that
\begin{equation}
  \mathrm{tr}\big(\mathbf J_{\mathrm{aux}}^{-1}\big)\le\epsilon, \mathbf A(\mathbf W,\boldsymbol\Phi)\succeq\mathbf J_{\mathrm {aux}}.
  \label{eq:aux_J_ineq}
\end{equation}
The matrix inequality~\eqref{eq:aux_J_ineq} can be recast via the Schur complement as a single LMI involving $\mathbf J_{\boldsymbol\theta\boldsymbol\theta}, \mathbf J_{\boldsymbol\theta\boldsymbol\alpha}, \mathbf J_{\boldsymbol\alpha\boldsymbol\alpha}$ and $\mathbf J_{\mathrm {aux}}$.
Indeed,~\eqref{eq:aux_J_ineq} is equivalent to $  \mathbf J_{\boldsymbol\theta\boldsymbol\theta}
  -
  \mathbf J_{\boldsymbol\theta\boldsymbol\alpha}
  \mathbf J_{\boldsymbol\alpha\boldsymbol\alpha}^{-1}
  \mathbf J_{\boldsymbol\alpha\boldsymbol\theta}
  -
  \mathbf J_{\mathrm {aux}}
  \succeq \mathbf 0$, which, by the Schur complement, is equivalent to~\cite{matrix}
\begin{equation}
  \mathbf M(\mathbf W,\boldsymbol\Phi,\mathbf J_{\mathrm {aux}})
  \triangleq
  \begin{bmatrix}
    \mathbf J_{\boldsymbol\theta\boldsymbol\theta}
    - \mathbf J_{\mathrm {aux}}
    &
    \mathbf J_{\boldsymbol\theta\boldsymbol\alpha}
    \\
    \mathbf J_{\boldsymbol\alpha\boldsymbol\theta}
    &
    \mathbf J_{\boldsymbol\alpha\boldsymbol\alpha}
  \end{bmatrix}
  \succeq \mathbf 0.
  \label{eq:Schur_CRB_LMI}
\end{equation}
Therefore, $\mathrm{tr}(\mathrm{CRB}_{\boldsymbol\theta}(\mathbf W,\boldsymbol\Phi))<\epsilon$ in~\eqref{prob:ISAC_joint} is equivalently
\begin{equation}
  \mathrm{tr}\big(\mathbf J_{\mathrm {aux}}^{-1}\big)\le\epsilon, \mathbf M(\mathbf W,\boldsymbol\Phi,\mathbf J_{\mathrm {aux}})
  \succeq \mathbf 0.
  \label{eq:CRB_M_LMI_final}
\end{equation}

We now show explicitly that $\mathbf J_{\boldsymbol\theta\boldsymbol\theta},\mathbf J_{\boldsymbol\theta\boldsymbol\alpha},\mathbf J_{\boldsymbol\alpha\boldsymbol\alpha}$ depend on $\mathbf W$ only through $\mathbf R_x=\mathbf W\mathbf W^*$. Using $\mathrm{vec}(\mathbf A\mathbf X)=(\mathbf X^{\mathrm T}\otimes\mathbf I)\mathrm{vec}(\mathbf A)$, we obtain
\begin{equation}
  \mathbf h(\boldsymbol\Phi,\boldsymbol\theta)
  =
  (\mathbf X^{\mathrm T}\otimes\mathbf I_{N_t})
  \mathrm{vec}\big(
    \mathbf H_t(\boldsymbol\Phi,\boldsymbol\theta)
  \big).
  \label{eq:h_vec_X}
\end{equation}
Similarly, for $i\in\{\mathrm B,\mathrm R\}$,
\begin{equation}
  \frac{\partial\mathbf h(\boldsymbol\Phi,\boldsymbol\theta)}
       {\partial\theta_i}
  =
  (\mathbf X^{\mathrm T}\otimes\mathbf I_{N_t})
  \mathrm{vec}\Big(
    \frac{\partial\mathbf H_t(\boldsymbol\Phi,\boldsymbol\theta)}
            {\partial\theta_i}
  \Big).
  \label{eq:dh_vec_X}
\end{equation}
Define the shorthand $  \mathbf p_i
  \triangleq
  \frac{\partial\mathbf h(\boldsymbol\Phi,\boldsymbol\theta)}
       {\partial\theta_i}$ and $
  \mathbf h
  \triangleq
  \mathbf h(\boldsymbol\Phi,\boldsymbol\theta)~(i\in\{\mathrm B,\mathrm R\})$. Using~\eqref{eq:dmu_dtheta}, the partial derivatives of $\boldsymbol\mu$ are
\begin{equation}
\label{pdmu}
  \frac{\partial\boldsymbol\mu}{\partial\theta_i}= \alpha_t \mathbf p_i,\frac{\partial\boldsymbol\mu}{\partial\alpha_{\mathrm R}}
  = \mathbf h,\frac{\partial\boldsymbol\mu}{\partial\alpha_{\mathrm I}} = j\mathbf h.
\end{equation}
Substituting into~\eqref{jta}, the FIM blocks take the form
\begin{equation}
\begin{aligned}
  [\mathbf J_{\boldsymbol\theta\boldsymbol\theta}]_{i,j}
  &=
  \frac{2}{\sigma_r^2}
  \Re\big\{
    \alpha_t^{*}\alpha_t
    \mathbf p_i^*\mathbf p_j
  \big\}
  =
  \frac{2|\alpha_t|^2}{\sigma_r^2}
  \Re\big\{
    \mathbf p_i^*\mathbf p_j
  \big\},\\
  \mathbf J_{\boldsymbol\alpha\boldsymbol\alpha}
  &=
  \frac{2}{\sigma_r^2}
  \Re\Big\{
    \big[\mathbf h~j\mathbf h\big]^{*}
    \big[\mathbf h~j\mathbf h\big]
  \Big\}
  =
  \frac{2\|\mathbf h\|_2^2}{\sigma_r^2}\mathbf I_2,
  \\
  [\mathbf J_{\boldsymbol\theta\boldsymbol\alpha}]_{i,1}
  &=
  \frac{2}{\sigma_r^2}
  \Re\big\{
    \alpha_t^{*}\mathbf p_i^*\mathbf h
  \big\}, [\mathbf J_{\boldsymbol\theta\boldsymbol\alpha}]_{i,2}
  =
  \frac{2}{\sigma_r^2}
  \Re\big\{
    j\alpha_t^{*}\mathbf p_i^*\mathbf h
  \big\}
  \label{eq:J_thetaalpha_I}
\end{aligned}
\end{equation}
for $i,j\in\{\mathrm B,\mathrm R\}\leftrightarrow\{1, 2\}$.

Next, we express $\mathbf p_i^*\mathbf p_j$ and $\mathbf p_i^*\mathbf h$ explicitly in terms of $\mathbf R_x = \mathbf W\mathbf W^*$. From~\eqref{eq:h_vec_X}-\eqref{eq:dh_vec_X}, we have
\begin{equation}
\begin{aligned}
  \mathbf p_i^*\mathbf p_j
  &=
  \mathrm{vec}\Big(
    \frac{\partial\mathbf H_t}{\partial\theta_i}
  \Big)^{*}
  (\mathbf X^{\mathrm{T}*}\mathbf X^{\mathrm T}\otimes\mathbf I_{N_t})
  \mathrm{vec}\Big(
    \frac{\partial\mathbf H_t}{\partial\theta_j}
  \Big) \\
  &=
  L
  \mathrm{vec}\Big(
    \frac{\partial\mathbf H_t}{\partial\theta_i}
  \Big)^{*}
  (\mathbf R_x^{\mathrm T}\otimes\mathbf I_{N_t})
  \mathrm{vec}\Big(
    \frac{\partial\mathbf H_t}{\partial\theta_j}
  \Big),
\end{aligned}
\label{eq:pipj_Rx}
\end{equation}
and
\begin{equation}
\begin{aligned}
  \mathbf p_i^*\mathbf h
  &=
  \mathrm{vec}\Big(
    \frac{\partial\mathbf H_t}{\partial\theta_i}
  \Big)^{*}
  (\mathbf X^{\mathrm{T}*}\mathbf X^{\mathrm T}\otimes\mathbf I_{N_t})
  \mathrm{vec}\big(
    \mathbf H_t
  \big) \\
  &=
  L
  \mathrm{vec}\Big(
    \frac{\partial\mathbf H_t}{\partial\theta_i}
  \Big)^{*}
  (\mathbf R_x^{\mathrm T}\otimes\mathbf I_{N_t})
  \mathrm{vec}\big(
    \mathbf H_t
  \big).
\end{aligned}
\label{eq:pih_Rx}
\end{equation}
Finally,
\begin{equation}
\begin{aligned}
  \|\mathbf h\|_2^2
  &=
  \mathrm{vec}(\mathbf H_t)^{*}
  (\mathbf X^{\mathrm{T}*}\mathbf X^{\mathrm T}\otimes\mathbf I_{N_t})
  \mathrm{vec}(\mathbf H_t) \\
  &=L  \mathrm{vec}(\mathbf H_t)^{*}  (\mathbf R_x^{\mathrm T}\otimes\mathbf I_{N_t})  \mathrm{vec}(\mathbf H_t).
\end{aligned}
\label{eq:h2_Rx}
\end{equation}
Equations~\eqref{eq:pipj_Rx}-\eqref{eq:h2_Rx} show that all entries of $\mathbf J_{\boldsymbol\theta\boldsymbol\theta}, \mathbf J_{\boldsymbol\theta\boldsymbol\alpha}$, and $\mathbf J_{\boldsymbol\alpha\boldsymbol\alpha}$ are affine functions of $\mathbf R_x = \mathbf W\mathbf W^*$.

In order to obtain a compact LMI representation of~\eqref{eq:CRB_M_LMI_final}, we define the followings:
\begin{equation}
\begin{aligned}
  F_1(\mathbf W,\boldsymbol\Phi)
  &\triangleq
  \mathbf p_{\mathrm B}^*\mathbf p_{\mathrm B},F_2(\mathbf W,\boldsymbol\Phi)\triangleq\mathbf p_{\mathrm B}^*\mathbf p_{\mathrm R},F_3(\mathbf W,\boldsymbol\Phi)
  \triangleq
  \alpha_t^{*}\mathbf p_{\mathrm B}^*\mathbf h,\\
  F_4(\mathbf W,\boldsymbol\Phi)&\triangleq\mathbf p_{\mathrm R}^*\mathbf p_{\mathrm R}, F_5(\mathbf W,\boldsymbol\Phi)\triangleq
  \alpha_t^{*}\mathbf p_{\mathrm R}^*\mathbf h, F_6(\mathbf W,\boldsymbol\Phi)
  \triangleq
    \|\mathbf h\|_2^2.
\label{deff1f6}
\end{aligned}
\end{equation}
By~\eqref{eq:pipj_Rx}-\eqref{eq:h2_Rx}, each $F_i$ is an affine function of $\mathbf R_x = \mathbf W\mathbf W^*$ and thus
a quadratic function of $\mathbf W$. Using these definitions together with~\eqref{eq:J_thetaalpha_I},
the FIM blocks can be written as
\begin{equation}
\begin{aligned}
 & \mathbf J_{\boldsymbol\theta\boldsymbol\theta}
  =
  \frac{2|\alpha_t|^2}{\sigma_r^2}\Re
  \begin{bmatrix}
    F_1 & F_2 \\
    F_2 & F_4
  \end{bmatrix},
  \mathbf J_{\boldsymbol\alpha\boldsymbol\alpha}
  =
  \frac{2F_6}{\sigma_r^2}\mathbf I_2,\\
 & \mathbf J_{\boldsymbol\theta\boldsymbol\alpha}
  =
  \frac{2}{\sigma_r^2}\Re
  \begin{bmatrix}
    \alpha_t^{*}\mathbf p_{\mathrm B}^*\mathbf h
    &
    j\alpha_t^{*}\mathbf p_{\mathrm B}^*\mathbf h\\
    \alpha_t^{*}\mathbf p_{\mathrm R}^*\mathbf h
    &
    j\alpha_t^{*}\mathbf p_{\mathrm R}^*\mathbf h
  \end{bmatrix}.
  \label{eq:J_alphaalpha_Fi}
  \end{aligned}
\end{equation}
For notational simplicity, we include them in the same parametrization and denote their collection by $\mathbf f = [f_1 \cdots f_6]^{\mathrm T}$ to equally constrain $F_i(\mathbf W,\boldsymbol\Phi)$, i.e., $f_i=F_i(\mathbf W,\boldsymbol\Phi)~(i=1, \cdots, 6)$. Substituting~\eqref{eq:J_alphaalpha_Fi} into~\eqref{eq:Schur_CRB_LMI}, $\mathbf M(\mathbf W,\boldsymbol\Phi,\mathbf J_{\mathrm {aux}})\succeq\mathbf0$ becomes a function of $\mathbf f$ and $\mathbf J_{\mathrm{aux}}$: $  \mathbf M(\mathbf f,\mathbf J_{\mathrm {aux}})\succeq \mathbf 0$, where $\mathbf M(\mathbf f,\mathbf J_{\mathrm {aux}})$ is an affine of $\mathbf f$ and $\mathbf J_{\mathrm {aux}}$, given by 
\begin{equation}
\mathbf M(\mathbf f,\mathbf J_{\mathrm{aux}})
=
\begin{bmatrix}
\frac{2|\alpha_t|^2}{\sigma_r^2}
\Re
\begin{bmatrix}
f_1 & f_2\\
f_2 & f_4
\end{bmatrix}-
\mathbf J_{\mathrm{aux}}
&
\frac{2}{\sigma_r^2}
\Re
\begin{bmatrix}
f_3 & j f_3\\
f_5 & j f_5
\end{bmatrix}
\\
\frac{2}{\sigma_r^2}
\Re
\begin{bmatrix}
f_3 & j f_3\\
f_5 & j f_5
\end{bmatrix}^{\mathrm{T}}
&
\frac{2}{\sigma_r^2} f_6 \mathbf I_2
\end{bmatrix}.
\label{eq:M_f_J_LMId}
\end{equation}
\subsection{Equivalent CRB-Constrained Optimization Problem}
Collecting the above results,~\eqref{prob:ISAC_joint} becomes
\begin{equation}
\label{prob:W_CRB_fixedPhi_full}
\begin{aligned}
&  \max_{\mathbf W,\boldsymbol\Phi, \mathbf r, \{\mathbf c_k\}, \mathbf J_{\mathrm {aux}},\mathbf f}~
   \mathcal F(\mathbf W,\boldsymbol\Phi,\mathbf r,\{\mathbf c_k\})\\
  \mathrm{s.t.}~
  &
  \mathrm{tr}\big(\mathbf J_{\mathrm {aux}}^{-1}\big)\le\epsilon, \mathbf J_{\mathrm {aux}}\succ0, \mathbf M(\mathbf f,\mathbf J_{\mathrm {aux}})
  \succeq \mathbf 0,
  \\
  &
  f_i = F_i(\mathbf W,\boldsymbol\Phi)~(i=1,\cdots,6),
  \\
  &
  \mathrm{tr}(\mathbf W\mathbf W^*)\le P_{\max}, |\boldsymbol\Phi_{n,n}| = 1~ (\forall n\in\{1,\cdots,N\}).
\end{aligned}
\end{equation}
where the CRB constraint has been fully encoded into the auxiliary $\mathbf J_{\mathrm {aux}}$ and $\mathbf f$ via~\eqref{eq:CRB_M_LMI_final} and~\eqref{eq:M_f_J_LMId}, together with $f_i=F_i(\mathbf W,\boldsymbol\Phi)$. 
The augmented Lagrangian formulation of~\eqref{prob:W_CRB_fixedPhi_full} with penalizing $\{ f_i = F_i(\mathbf W,\boldsymbol\Phi)\}$ is formulated as
\begin{equation}
\begin{aligned}
&\min_{\substack{\mathbf W,\boldsymbol\Phi,\mathbf r,\{\mathbf c_k\},\\ \mathbf J_{\mathrm{aux}},\mathbf f}}
 -\mathcal F(\mathbf W,\boldsymbol\Phi,\mathbf r,\{\mathbf c_k\})\\
 &~~~~~~~~~~~~~+ \frac{1}{2\rho_1}\sum_{i=1}^{6}\left|F_i(\mathbf W,\boldsymbol\Phi)- f_i+ \rho_1 \zeta_i\right|^2\\
  &\mathrm{s.t.}~\mathrm{tr}\big(\mathbf J_{\mathrm {aux}}^{-1}\big)\le\epsilon, \mathbf J_{\mathrm {aux}}\succ0, \mathbf M(\mathbf f,\mathbf J_{\mathrm {aux}})
  \succeq \mathbf 0,\\
  &\mathrm{tr}(\mathbf W\mathbf W^*)\le P_{\max}, |\boldsymbol\Phi_{n,n}| = 1~ (\forall n\in\{1,\cdots,N\}),
\end{aligned}
\label{eq:AL_final}
\end{equation}
where $\boldsymbol\zeta=[\zeta_1 \cdots \zeta_6]^{\mathrm T}$ denotes the vector of dual variables and $\rho_1>0$ is the associated penalty parameter. Based on this formulation, the variables can be updated in an alternating fashion using a BCD procedure.
\subsection{Updating Blocks}\label{ubub}
\subsubsection{Update $\mathbf r$ and $\{\mathbf c_k\}$}
Fixing the remaining variables, the optimal $\mathbf r$ and $\{\mathbf c_k\}$ are directly obtained from~\eqref{rkmax} and~\eqref{ckmax}.
\subsubsection{Update of $\mathbf J_{\mathrm{aux}}$ and $\mathbf f$}
With all other variables fixed, the update of $\mathbf J_{\mathrm{aux}}$ and $\mathbf f$ is achieved by solving
\begin{equation}
\label{jfsdp_para}
\begin{aligned}
&\min_{\mathbf J_{\mathrm{aux}}, \mathbf f}~\sum_{i=1}^{6}\big|F_i(\mathbf W,\boldsymbol\Phi)- f_i+ \rho_1\zeta_i\big|^2 \\
&\text{s.t.}~\mathrm{tr}\left(\mathbf J_{\mathrm{aux}}^{-1}\right)\le\epsilon, \mathbf J_{\mathrm{aux}}\succ 0, ~\mathbf M(\mathbf f,\mathbf J_{\mathrm{aux}})\succeq 0,
\end{aligned}
\end{equation}
which constitutes a convex semidefinite program (SDP) and can be efficiently solved using standard SDP solvers~\cite{boyd}.
\subsubsection{Update $\mathbf W$}
With other variables fixed,~\eqref{eq:AL_final} reduces to:
\begin{equation}
\begin{aligned}
\min_{\mathbf W}~
&\|\mathbf B\mathbf w\|_2^2-\Re\{\mathbf a^{*}\mathbf w\}+\frac{1}{2\rho_1}\sum_{i=1}^{6}\big|F_i(\mathbf W,\boldsymbol\Phi)- f_i+ \rho_1\zeta_i\big|^2\\
\mathrm{s.t.}~&\mathrm{tr}(\mathbf W\mathbf W^*)\le P_{\max},
\end{aligned}
\label{eq:W_subprob_raw}
\end{equation}
where the first two terms come from~\eqref{eq:FP_compact_w_vector_final}. Using $\mathrm{vec}(\mathbf A)^{*}
  (\mathbf R^{\mathrm T}\otimes\mathbf I)
  \mathrm{vec}(\mathbf B)
  =
  \mathrm{tr}\left(
    \mathbf R\mathbf A^{*}\mathbf B
  \right)$, we obtain the explicit forms from~\eqref{deff1f6}, which is organized in~\eqref{eq:Fi_tr_repr}. 
\begin{figure*}
\begin{equation}
\begin{aligned}
F_1(\mathbf W,\boldsymbol\Phi)
  &= 
  L\mathrm{tr}\left(\mathbf R_x
    \frac{\partial\mathbf H_t^{*}}{\partial\theta_{\mathrm B}}
    \frac{\partial\mathbf H_t}{\partial\theta_{\mathrm B}}
  \right), F_2(\mathbf W,\boldsymbol\Phi)= 
  L\mathrm{tr}\left(\mathbf R_x
    \frac{\partial\mathbf H_t^{*}}{\partial\theta_{\mathrm B}}
    \frac{\partial\mathbf H_t}{\partial\theta_{\mathrm R}}
  \right), F_3(\mathbf W,\boldsymbol\Phi)= 
  L\mathrm{tr}\left(\mathbf R_x
    \alpha_t^{*}
    \frac{\partial\mathbf H_t^{*}}{\partial\theta_{\mathrm B}}
    \mathbf H_t
  \right),\\
F_4(\mathbf W,\boldsymbol\Phi)
  &= 
  L\mathrm{tr}\left(\mathbf R_x
    \frac{\partial\mathbf H_t^{*}}{\partial\theta_{\mathrm R}}
    \frac{\partial\mathbf H_t}{\partial\theta_{\mathrm R}}
  \right), 
F_5(\mathbf W,\boldsymbol\Phi)
  = 
  L\mathrm{tr}\left(\mathbf R_x
    \alpha_t^{*}
    \frac{\partial\mathbf H_t^{*}}{\partial\theta_{\mathrm R}}
    \mathbf H_t
  \right), F_6(\mathbf W,\boldsymbol\Phi)
  = 
  L\mathrm{tr}\left(\mathbf R_x
    \mathbf H_t^{*}\mathbf H_t
  \right).
\end{aligned}
\label{eq:Fi_tr_repr}
\end{equation}
\hrule
\end{figure*}
Thus, we can define $\mathbf A_i(\boldsymbol\Phi)\in\mathbb C^{N_t\times N_t}$ such that
\begin{equation}
F_i(\mathbf W,\boldsymbol\Phi)
=
\mathrm{tr}\left(
  \mathbf A_i(\boldsymbol\Phi)\mathbf W\mathbf W^*
\right)
\label{eq:Fi_quad_form}
\end{equation}
with
\begin{equation}
\begin{aligned}
\mathbf A_1 &= 
L\frac{\partial\mathbf H_t^{*}}{\partial\theta_{\mathrm B}}
 \frac{\partial\mathbf H_t}{\partial\theta_{\mathrm B}},
\mathbf A_2 =
L\frac{\partial\mathbf H_t^{*}}{\partial\theta_{\mathrm B}}
 \frac{\partial\mathbf H_t}{\partial\theta_{\mathrm R}},\mathbf A_3 = 
L\alpha_t^{*}
 \frac{\partial\mathbf H_t^{*}}{\partial\theta_{\mathrm B}}
 \mathbf H_t,\\
 \mathbf A_4 &=
L\frac{\partial\mathbf H_t^{*}}{\partial\theta_{\mathrm R}}
 \frac{\partial\mathbf H_t}{\partial\theta_{\mathrm R}},
\mathbf A_5 = 
L\alpha_t^{*}
 \frac{\partial\mathbf H_t^{*}}{\partial\theta_{\mathrm R}}
 \mathbf H_t,
\mathbf A_6 =
L\mathbf H_t^{*}\mathbf H_t .
\end{aligned}
\label{eq:Ai_explicit}
\end{equation}
Using $\mathbf w=\mathrm{vec}(\mathbf W)$ and $\mathrm{tr}(\mathbf A\mathbf W\mathbf W^*)=
\mathbf w^{*}(\mathbf I_K\otimes\mathbf A)\mathbf w$, we obtain the compact quadratic form
\begin{equation}
F_i(\mathbf W,\boldsymbol\Phi)=\mathbf w^{*}\mathbf C_i\mathbf w, \mathbf C_i\triangleq \mathbf I_K\otimes\mathbf A_i(\boldsymbol\Phi).
\label{eq:Fi_Ci_compact}
\end{equation}
Hence, by letting $d_i\triangleq - f_i + \rho_1 \zeta_i$ and substituting~\eqref{eq:Fi_Ci_compact} into~\eqref{eq:W_subprob_raw}, it becomes
\begin{equation}
\begin{aligned}
\min_{\mathbf w}~
&\|\mathbf B\mathbf w\|_2^2
-
\Re\{\mathbf a^{*}\mathbf w\}
+
\frac{1}{2\rho_1}
\sum_{i=1}^{6}
\left|
\mathbf w^{*}\mathbf C_i\mathbf w + d_i
\right|^2\\
\mathrm{s.t.}~&
\|\mathbf w\|_2^2\le P_{\max}.
\end{aligned}
\label{eq:W_quartic_form}
\end{equation}
Problem~\eqref{eq:W_quartic_form} is still nonconvex since the last term in objective is quartic in $\mathbf w$.  To tackle this, we adopt an MM strategy~\cite{Liu_RIS_ISAC_CRB}, where at each iteration we replace the quartic penalty by a tight quadratic surrogate around the current iterate.

Let $t$ denote the iteration index and $\mathbf w^{(t)}=\operatorname{vec}(\mathbf W^{(t)})$ be the precoder vector with precoder $\mathbf W^{(t)}$ at iteration $t$. For brevity, define $z_i(\mathbf w)\triangleq \mathbf w^{*}\mathbf C_i\mathbf w + d_i~(i=1,\cdots,6)$. Each quartic term can then be expanded as
\begin{equation}
\label{eq:quartic_expand}
\begin{aligned}
  |z_i(\mathbf w)|^2
  &= |\mathbf w^{*}\mathbf C_i\mathbf w|^2
     + 2\Re \left\{d_i^{*}\mathbf w^{*}\mathbf C_i\mathbf w\right\}
     + |d_i|^2.
\end{aligned}
\end{equation}
The constant $|d_i|^2$ does not affect the minimizer and will be omitted in the sequel. Define $\bar{\mathbf C}_i\triangleq d_i^{*}\mathbf C_i + d_i\mathbf C_i^{*}$, which is Hermitian, so that the second term of~\eqref{eq:quartic_expand} can be written as
\begin{equation}
  2\Re\left\{d_i^{*}\mathbf w^{*}\mathbf C_i\mathbf w\right\}
  = \mathbf w^{*}\bar{\mathbf C}_i\mathbf w.
  \label{eq:second_term_quad}
\end{equation}
Let $\vartheta_{t,i}\ge\lambda_{\max}(\bar{\mathbf C}_i)$ with largest eigenvalue $\lambda_{\max}(\cdot)$. Then $(\mathbf w - \mathbf w^{(t)})^{*}(\vartheta\mathbf I - \bar{\mathbf C}_i)(\mathbf w - \mathbf w^{(t)}) \ge 0$ holds, and we obtain
\begin{equation}
\label{eq:mm_linearize_quad}
\mathbf w^{*}\bar{\mathbf C}_i\mathbf w
\le 2\Re \left\{
  \mathbf w^{*}(\bar{\mathbf C}_i-\vartheta_{t,i}\mathbf I)\mathbf w^{(t)}
\right\}
+\varpi_{1,i}^{(t)},
\end{equation}
where $ \varpi_{1,i}^{(t)}
  \triangleq
  \vartheta_{t,i}P_{\max}
  +\mathbf w^{(t)*}
  \big(
    \vartheta_{t,i}\mathbf I-\bar{\mathbf C}_i
  \big)
  \mathbf w^{(t)}$ is a constant independent of $\mathbf w$, and we have used $\|\mathbf w\|_2^2\le P_{\max}$ to upper-bound the term involving $\|\mathbf w\|_2^2$. The right-hand side of~\eqref{eq:mm_linearize_quad} is affine in $\mathbf w$ and  coincides with $\mathbf w^{*}\bar{\mathbf C}_i\mathbf w$ at $\mathbf w=\mathbf w^{(t)}$, thus serving as a valid MM surrogate.

To majorize the first term of~\eqref{eq:quartic_expand}, we may equivalently express it using the Hermitian part of $\bar{\mathbf A}_i$:
\begin{equation}
\label{eq:Ai_Hermitian}
|\mathbf w^{*}\mathbf C_i\mathbf w|^2
=
\bar{\mathbf w}^{*}\widehat{\mathbf A}_i\bar{\mathbf w}, \widehat{\mathbf A}_i
\triangleq
\frac{\bar{\mathbf A}_i+\bar{\mathbf A}_i^{*}}{2},
\end{equation}
where $\bar{\mathbf A}_i=\mathbf C_i^{\mathrm T}\otimes \mathbf C_i^{*}$ and $\bar{\mathbf w}=\mathrm{vec}(\mathbf w\mathbf w^{*})$. By construction, $\widehat{\mathbf A}_i$ is Hermitian. Let $\vartheta_{b,i}\ge \lambda_{\max}(\widehat{\mathbf A}_i)$.
Since 
$\vartheta_{b,i}\mathbf I - \widehat{\mathbf A}_i\succeq \mathbf 0$,
it holds that $(\bar{\mathbf w}-\bar{\mathbf w}^{(t)})^{*}
\big(\vartheta_{b,i}\mathbf I-\widehat{\mathbf A}_i\big)
(\bar{\mathbf w}-\bar{\mathbf w}^{(t)})\ge 0$, where $\bar{\mathbf w}^{(t)}=\mathrm{vec}(\mathbf w^{(t)}\mathbf w^{(t)*})$ be $\bar{\mathbf w}$ at iteration $t$. Similar to~\eqref{eq:mm_linearize_quad}, expanding it yields
\begin{equation}
\label{eq:quartic_majorized_final}
\bar{\mathbf w}^{*}\widehat{\mathbf A}_i\bar{\mathbf w}
\le
2\Re\left\{
\bar{\mathbf w}^{*}
\big(\widehat{\mathbf A}_i-\vartheta_{b,i}\mathbf I\big)
\bar{\mathbf w}^{(t)}
\right\}
+
\varpi_{2,i}^{(t)},
\end{equation}
where $\varpi_{2,i}^{(t)}
\triangleq
\bar{\mathbf w}^{(t) *}
\big(\vartheta_{b,i}\mathbf I-\widehat{\mathbf A}_i\big)
\bar{\mathbf w}^{(t)}+\vartheta_{b,i} P_{\max}^2$ is independent of $\bar{\mathbf w}$ and can be treated as a constant, and we used
\begin{equation}
\label{wsqp}
\bar{\mathbf w}=\mathrm{vec}(\mathbf w \mathbf w^*)
\rightarrow
\|\bar{\mathbf w}\|_2^2
= \|\mathbf w\mathbf w^*\|_F^2
= \|\mathbf w\|_2^4\le P_{\max}^2.
\end{equation}
Thereafter, we expand the real-valued bilinear term as
\begin{equation}
\begin{aligned}
&\Re\left\{ \bar{\mathbf w}^{(t)*}
\big(\bar{\mathbf A}_i+\bar{\mathbf A}_i^*-2\vartheta_{b,i}\mathbf I\big)
\bar{\mathbf w}
\right\}
+
\varpi_{2,i}^{(t)}
\\
&=
\Re\{\bar{\mathbf w}^{(t)*}\bar{\mathbf A}_i\bar{\mathbf w}\}
+
\Re\{\bar{\mathbf w}^{(t)*}\bar{\mathbf A}_i^*\bar{\mathbf w}\}\\
&~~~~-2\vartheta_{b,i}\Re\{\bar{\mathbf w}^{(t)*}\bar{\mathbf w}\}
+
\varpi_{2,i}^{(t)}.
\end{aligned}
\label{eq:47c_expand}
\end{equation}
Therein, each term in~\eqref{eq:47c_expand} admits the closed-form expressions:
\begin{equation}
\begin{aligned}
\label{wve}
\Re\{\bar{\mathbf w}^{(t)*}\bar{\mathbf A}_i\bar{\mathbf w}\}
&=\big|\mathbf w^*\mathrm{vec}(\mathbf A_i\mathbf W^{(t)})\big|^{2},\\
\Re\{\bar{\mathbf w}^{(t)*}\bar{\mathbf A}_i^*\bar{\mathbf w}\}
&=
\big|\mathbf w^*\mathrm{vec}(\mathbf A_i^*\mathbf W^{(t)})\big|^{2},\\
\Re\{\bar{\mathbf w}^{(t)*}\bar{\mathbf w}\}
&=
\mathbf w^*\mathbf w^{(t)}\mathbf w^{(t)*}\mathbf w.
\end{aligned}
\end{equation}
To show this, we first derive the closed-form of $\Re\{\bar{\mathbf w}^{(t)*}\bar{\mathbf A}_i\bar{\mathbf w}\}$. Using $\mathrm{vec}(\mathbf X)^*(\mathbf B\otimes\mathbf C)\mathrm{vec}(\mathbf Y)=\mathrm{tr}\left(\mathbf X^*\mathbf C\mathbf Y\mathbf B^{\mathrm{T}}\right)$, we obtain
\begin{equation}
\begin{aligned}
\bar{\mathbf w}^{(t)*}\bar{\mathbf A}_i\bar{\mathbf w}
&=
\mathrm{vec}(\mathbf w^{(t)}\mathbf w^{(t)*})^*
(\mathbf C_i^{\mathrm{T}}\otimes\mathbf C_i^{*})
\mathrm{vec}(\mathbf w\mathbf w^*) \\
&=\mathrm{tr}\left(
(\mathbf w^{(t)}\mathbf w^{(t)*})^*
\mathbf C_i^{*}
(\mathbf w\mathbf w^*)
\mathbf C_i
\right) \\
&=\mathrm{tr}\left(\mathbf w^{(t)*}\mathbf C_i^{*}\mathbf w
\mathbf w^*\mathbf C_i\mathbf w^{(t)}\right).
\label{eq:first_trace_2}
\end{aligned}
\end{equation}
The scalar inside the trace is rank-one, hence
\begin{equation}
\label{eq:first_rank1}
\bar{\mathbf w}^{(t)*}\bar{\mathbf A}_i\bar{\mathbf w}
=
\big(\mathbf w^*\mathbf C_i\mathbf w^{(t)}\big)
\big(\mathbf w^{(t)*}\mathbf C_i^*\mathbf w\big)
=
\left|\mathbf w^*\mathbf C_i\mathbf w^{(t)}\right|^{2}.
\end{equation}
Since the right-hand side is already real and nonnegative:
\begin{equation}
\label{eq:first_final}
\Re\{\bar{\mathbf w}^{(t)*}\bar{\mathbf A}_i\bar{\mathbf w}\}
=
\left|\mathbf w^*\mathbf C_i\mathbf w^{(t)}\right|^{2}=\big|\mathbf w^*\mathrm{vec}(\mathbf A_i\mathbf W^{(t)})\big|^{2}
\end{equation}
since $\mathbf C_i \mathbf w^{(t)}=(\mathbf I_K\otimes \mathbf A_i)\mathrm{vec}(\mathbf W^{(t)})=\mathrm{vec}(\mathbf A_i \mathbf W^{(t)})$ using $(\mathbf I\otimes \mathbf A)\mathrm{vec}(\mathbf X)=\mathrm{vec}(\mathbf A\mathbf X)$, and so is $\Re\{\bar{\mathbf w}^{(t)*}\bar{\mathbf A}_i^*\bar{\mathbf w}\}$.

We then derive the closed-form of $\Re\{\bar{\mathbf w}^{(t)*}\bar{\mathbf w}\}$. By using $\mathrm{vec}(\mathbf X)^*\mathrm{vec}(\mathbf Y)=\mathrm{tr}(\mathbf X^*\mathbf Y)$, we first have
\begin{equation}
\begin{aligned}
\bar{\mathbf w}^{(t)*}\bar{\mathbf w}
&=
\mathrm{vec}(\mathbf w^{(t)}\mathbf w^{(t)*})^*
\mathrm{vec}(\mathbf w\mathbf w^*) \\
&=
\mathrm{tr}\left(
(\mathbf w^{(t)}\mathbf w^{(t)*})^*(\mathbf w\mathbf w^*)
\right) \\
&=\mathbf w^{(t)*}\mathbf w \mathbf w^*\mathbf w^{(t)}=\mathbf w^*\mathbf w^{(t)}\mathbf w^{(t)*}\mathbf w\in\mathbb R.
\label{eq:third_rank1}
\end{aligned}
\end{equation}
Thus it becomes $\Re\{\bar{\mathbf w}^{(t)*}\bar{\mathbf w}\}=\mathbf w^*\mathbf w^{(t)}\mathbf w^{(t)*}\mathbf w$. Hence, substituting these into~\eqref{eq:47c_expand} yields the closed-form surrogate
\begin{equation}
\begin{aligned}
&\Re\left\{ \bar{\mathbf w}^{(t)*}
\big(\bar{\mathbf A}_i+\bar{\mathbf A}_i^*-2\vartheta_{b,i}\mathbf I\big)
\bar{\mathbf w}
\right\}
+
\varpi_{2,i}^{(t)}
\\
&=
\big|\mathbf w^*\mathrm{vec}(\mathbf A_i\mathbf W^{(t)})\big|^{2}
+
\big|\mathbf w^*\mathrm{vec}(\mathbf A_i^*\mathbf W^{(t)})\big|^{2}
\\
&~~~~
-2\vartheta_{b,i}\mathbf w^*\mathbf w^{(t)}\mathbf w^{(t)*}\mathbf w
+\varpi_{2,i}^{(t)}.
\end{aligned}
\label{eq:47d_final}
\end{equation}
We see that only the third term of the right-hand side of~\eqref{eq:47d_final} is non-convex. To majorize this term, we apply the first-order Taylor expansion at $\mathbf w^{(t)}$, which admits the global upper-bound
\begin{equation}
\mathbf w^*\mathbf w^{(t)}\mathbf w^{(t)*}\mathbf w\ge \|\mathbf w^{(t)}\|_2^4
+
2\Re\left\{
  \|\mathbf w^{(t)}\|_{2}^{2}\mathbf w^{(t)*}
  (\mathbf w-\mathbf w^{(t)})
\right\},
\label{eq:gi_taylor}
\end{equation}
Substituting the results obtained in~\eqref{eq:mm_linearize_quad},~\eqref{eq:47c_expand} and~\eqref{eq:gi_taylor} into~\eqref{eq:W_quartic_form}, by letting $\mathbf a_i^{(t)} \triangleq 2(\bar{\mathbf C}_i-\vartheta_{t,i}\mathbf I)\mathbf w^{(t)}-4\vartheta_{b,i}\|\mathbf w^{(t)}\|_{2}^{2}\mathbf w^{(t)}$, the sub-problem for updating $\mathbf w$ becomes:
\begin{equation}
\begin{aligned}
\min_{\mathbf w}~&\|\mathbf B\mathbf w\|_{2}^{2}-\Re\{\mathbf a^{*}\mathbf w\}+\frac{1}{2\rho_{1}}\sum_{i=1}^{6}\Big( \big|\mathbf w^*\mathrm{vec}(\mathbf A_i\mathbf W^{(t)})\big|^{2}\\
&+\big|\mathbf w^*\mathrm{vec}(\mathbf A_i^*\mathbf W^{(t)})\big|^{2}+\Re\{\mathbf a_{i}^{(t)*}\mathbf w \}\Big)
\label{eq:w_subProb_final}
\\
\text{s.t.}~&\|\mathbf w\|_{2}^{2}\le P_{\max}.
\end{aligned}
\end{equation}
Problem~\eqref{eq:w_subProb_final} is a convex quadratic program (QP), which can be efficiently solved using standard convex solvers~\cite{boyd}. The resulting $\mathbf w(\leftrightarrow\mathbf W)$ is then taken as the updated $\mathbf W^{(t+1)}$
\subsubsection{Update of $\boldsymbol\Phi$}
Fixing other variables and by~\eqref{eq:FP_compact_phi_vector_final}, the sub-problem of solving for $\boldsymbol\Phi(\leftrightarrow \boldsymbol\phi)$ is formulated as
\begin{equation}
\begin{aligned}
\min_{\boldsymbol\phi}~
&  \boldsymbol\phi^{*}\mathbf D \boldsymbol\phi
  -
  \Re\{\mathbf g^{*}\boldsymbol\phi\}
+
\frac{1}{2\rho_1}
\sum_{i=1}^{6}
\big|
F_i(\mathbf W,\boldsymbol\Phi)
+d_i\big|^2\\
\mathrm{s.t.}~&|\phi_n|=1~(\forall n),
\end{aligned}
\label{eq:phi_raw_subprob_final}
\end{equation}
where $\phi_n$ is the $n$th element of $\boldsymbol\phi$. Problem~\eqref{eq:phi_raw_subprob_final} is nonconvex due to both (i)~the unit-modulus constraints and (ii)~the quartic dependence of $F_i(\mathbf W,\boldsymbol\Phi)$ on $\boldsymbol\phi$.
To derive a tractable update, we again adopt a MM approach. We denote $\mathbf d_i^{(t)}
    \triangleq
    \nabla_{\boldsymbol\phi}
    F_i(\mathbf W,\boldsymbol\Phi^{(t)})
    \in\mathbb C^{N}$ with the Wirtinger gradient $\nabla_{\boldsymbol\phi}$ at iteration $t$ with corresponding $\boldsymbol\Phi^{(t)}$, and let the residual in~\eqref{eq:phi_raw_subprob_final} be $
    e_i^{(t)}
    \triangleq
    F_i(\mathbf W,\boldsymbol\Phi^{(t)}) +d_i$. The explicit form of $\mathbf d_i^{(t)}$ is given in Appendix~A. Using the first-order Taylor approximation and Lipschitz continuity of $\nabla_{\boldsymbol\phi}F_i$ with Lipschitz constant $\tau_i$:
\begin{equation}
\begin{aligned}
F_i(\mathbf W,\boldsymbol\Phi)
\le&
F_i(\mathbf W,\boldsymbol\Phi^{(t)})
+
2\Re\left\{\mathbf d_i^{(t)*}
    (\boldsymbol\phi-\boldsymbol\phi^{(t)})
\right\}\\
&+\frac{\tau_i}{2}\|\boldsymbol\phi-\boldsymbol\phi^{(t)}\|_2^{2},
\label{eq:Fi_MM_majorizer}
\end{aligned}
\end{equation}
where $\boldsymbol\Phi^{(t)}=\mathrm{diag}(\boldsymbol\phi^{(t)})$. Substituting~\eqref{eq:Fi_MM_majorizer} into the squared residual in~\eqref{eq:phi_raw_subprob_final} yields the quadratic upper-bound
\begin{equation}
\label{qubb}
|F_i(\mathbf W,\boldsymbol\Phi)+d_i|^2
\le
\boldsymbol\phi^{*}\mathbf Q_i^{(t)}\boldsymbol\phi
-
2\Re\left\{\mathbf q_i^{(t)*}\boldsymbol\phi
\right\}
+
\kappa_i^{(t)},
\end{equation}
where $\kappa_i^{(t)}$ is a term independent of $\boldsymbol\phi$, $\mathbf Q_i^{(t)}
\triangleq
2\mathbf d_i^{(t)}\mathbf d_i^{(t)*}+\tau_i\mathbf I$, and $\mathbf q_i^{(t)}
\triangleq
2e_i^{(t)}\mathbf d_i^{(t)}+\tau_i \boldsymbol\phi^{(t)}$. Discarding all constant terms, the MM surrogate problem of~\eqref{eq:phi_raw_subprob_final} becomes
\begin{equation}
\begin{aligned}
\min_{\boldsymbol\phi}~
\boldsymbol\phi^{*}\widehat{\mathbf D}^{(t)}\boldsymbol\phi
-
2\Re\left\{\widehat{\mathbf g}^{(t)*}\boldsymbol\phi
\right\}~\mathrm{s.t.}~|\phi_n|=1~(\forall n),
\end{aligned}
\label{eq:phi_MM_quadratic_final}
\end{equation}
where $\widehat{\mathbf D}^{(t)}
\triangleq
\mathbf D
+
\frac{1}{2\rho_1}\sum_{i=1}^{6}\mathbf Q_i^{(t)}$ and $\widehat{\mathbf g}^{(t)}
\triangleq\mathbf g+\frac{1}{2\rho_1}\sum_{i=1}^{6}\mathbf q_i^{(t)}$. We now solve~\eqref{eq:phi_MM_quadratic_final} by means of an ADMM.
Introduce an auxiliary variable $\mathbf z\in\mathbb C^{N}$ that carries the unit-modulus constraint, and enforce $\boldsymbol\phi=\mathbf z$ via a consensus constraint. Then~\eqref{eq:phi_MM_quadratic_final} is equivalently written as
\begin{equation}
\begin{aligned}
\min_{\boldsymbol\phi,\mathbf z}~
&
\boldsymbol\phi^{*}\widehat{\mathbf D}^{(t)}\boldsymbol\phi
-2\Re\{ \widehat{\mathbf g}^{(t)*}\boldsymbol\phi\}
+
\mathcal I_{\mathcal U}(\mathbf z)~\mathrm{s.t.}~\boldsymbol\phi-\mathbf z=\mathbf 0,
\end{aligned}
\label{eq:phi_ADMM_split}
\end{equation}
where $\mathcal U \triangleq \{\mathbf z\in\mathbb C^{N}:|z_n|=1~(\forall n)\}$, and $\mathcal I_{\mathcal U}(\cdot)$ denotes the indicator function of $\mathcal U$:
\begin{equation}
\label{indicu}
  \mathcal I_{\mathcal U}(\mathbf z)= 0~(\mathbf z\in\mathcal U),~\infty~(\text{otherwise}).
\end{equation}
Let $\boldsymbol\lambda\in\mathbb C^{N}$ be the dual variable associated with the constraint $\boldsymbol\phi-\mathbf z=\mathbf 0$, and let $\rho_{\phi}>0$ be the ADMM penalty parameter. The (unscaled) augmented Lagrangian of~\eqref{eq:phi_ADMM_split} is
\begin{equation}
\begin{aligned}
\mathcal L_{\rho_{\phi}}(\boldsymbol\phi,\mathbf z,\boldsymbol\lambda)
&=
\boldsymbol\phi^{*}\widehat{\mathbf D}^{(t)}\boldsymbol\phi
-
2\Re\{\widehat{\mathbf g}^{(t)*}\boldsymbol\phi\}
+\mathcal I_{\mathcal U}(\mathbf z)
\\
&~~~~+
\Re\big\{\boldsymbol\lambda^{*}(\boldsymbol\phi-\mathbf z)\big\}+\frac{\rho_{\phi}}{2}\|\boldsymbol\phi-\mathbf z\|_2^{2}.
\end{aligned}
\label{eq:phi_augLag}
\end{equation}
Herein, ADMM proceeds by cyclically minimizing $\mathcal L_{\rho_{\phi}}$ with respect to $\boldsymbol\phi$ and $\mathbf z$, followed by a dual update of $\boldsymbol\lambda$.
\paragraph*{$\boldsymbol\phi$\textbf{-update}}
Given $(\mathbf z^{[r]},\boldsymbol\lambda^{[r]})$ at iteration $r$ for fixed $t$, the $\boldsymbol\phi$-subproblem is
\begin{equation}
\begin{aligned}
\boldsymbol\phi^{[r+1]}
=
\arg\min_{\boldsymbol\phi}~& \boldsymbol\phi^{*}\widehat{\mathbf D}^{(t)}\boldsymbol\phi
-
2\Re\{\widehat{\mathbf g}^{(t)*}\boldsymbol\phi\}
\\
&+\Re\big\{\boldsymbol\lambda^{[r]*}(\boldsymbol\phi-\mathbf z^{[r]})\big\}
+\frac{\rho_{\phi}}{2}\|\boldsymbol\phi-\mathbf z^{[r]}\|_2^{2}.
\end{aligned}
\label{eq:phi_subprob_ADMM}
\end{equation}
Discarding the terms independent of $\boldsymbol\phi$, we obtain the strictly convex quadratic problem
\begin{equation}
\label{scqp}
\begin{aligned}
\min_{\boldsymbol\phi}~
&
\boldsymbol\phi^{*}\widehat{\mathbf D}^{(t)}\boldsymbol\phi
-
2\Re\{\widehat{\mathbf g}^{(t)*}\boldsymbol\phi\}
+
\Re\{\boldsymbol\lambda^{[r]*}\boldsymbol\phi\}
+
\frac{\rho_{\phi}}{2}
\|\boldsymbol\phi-\mathbf z^{[r]}\|_2^{2}.
\end{aligned}
\end{equation}
The first-order optimality condition of~\eqref{scqp} yields
\begin{equation}
\label{foc22}
  \big(2\widehat{\mathbf D}^{(t)}+\rho_{\phi}\mathbf I\big)\boldsymbol\phi
  =
  2\widehat{\mathbf g}^{(t)}
  -\boldsymbol\lambda^{[r]}
  +\rho_{\phi}\mathbf z^{[r]}.
\end{equation}
Since $\widehat{\mathbf D}^{(t)}\succeq\mathbf 0$ and $\rho_{\phi}>0$, $2\widehat{\mathbf D}^{(t)}+\rho_{\phi}\mathbf I\succ\mathbf 0$ holds:
\begin{equation}
\boldsymbol\phi^{[r+1]}
=
\big(2\widehat{\mathbf D}^{(t)}+\rho_{\phi}\mathbf I\big)^{-1}
\Big(
  2\widehat{\mathbf g}^{(t)}
  -\boldsymbol\lambda^{[r]}
  +\rho_{\phi}\mathbf z^{[r]}
\Big).
\label{eq:phi_ADMM_update_closed}
\end{equation}
Note that $\widehat{\mathbf D}^{(t)}$ is fixed within the iteration $r$, hence the inverse of $2\widehat{\mathbf D}^{(t)}+\rho_{\phi}\mathbf I$ in~\eqref{eq:phi_ADMM_update_closed} can be precomputed and reused across inner ADMM iterations.

\paragraph*{$\mathbf z$\textbf{-update}}
Given $\boldsymbol\phi^{[r+1]}$ and $\boldsymbol\lambda^{[r]}$, the $\mathbf z$-subproblem is
\begin{equation}
\begin{aligned}
\mathbf z^{[r+1]}
&=
\arg\min_{\mathbf z}~\mathcal I_{\mathcal U}(\mathbf z)
-
\Re\{\boldsymbol\lambda^{[r]*}\mathbf z\}
+
\frac{\rho_{\phi}}{2}\|
\boldsymbol\phi^{[r+1]}-\mathbf z
\|_2^{2}.
\end{aligned}
\label{eq:z_subprob_ADMM_raw}
\end{equation}
By completing the square, this is equivalent to
\begin{equation}
\mathbf z^{[r+1]}
=
\arg\min_{\mathbf z\in\mathcal U}
\left\|
  \mathbf z
  -
  \Big(
    \boldsymbol\phi^{[r+1]}
    +
    \frac{1}{\rho_{\phi}}\boldsymbol\lambda^{[r]}
  \Big)
\right\|_2^{2},
\end{equation}
i.e., projecting $\boldsymbol\phi^{[r+1]}+\frac{1}{\rho_{\phi}}\boldsymbol\lambda^{[r]}$ onto the unit-modulus set $\mathcal U$, which admits the closed-form with element-wise exponent:
\begin{equation}
\mathbf z^{[r+1]}
  =  e^{   j\angle\Big(\boldsymbol\phi^{[r+1]} +    \frac{1}{\rho_{\phi}}\boldsymbol\lambda^{[r]}\Big)},
\label{eq:z_ADMM_update}
\end{equation}
\paragraph*{$\boldsymbol\lambda$\textbf{-update}}
Finally, $\boldsymbol\lambda$ is updated as
\begin{equation}
\boldsymbol\lambda^{[r+1]}
=
\boldsymbol\lambda^{[r]}
+
\rho_{\phi}\big(
  \boldsymbol\phi^{[r+1]}-\mathbf z^{[r+1]}
\big).
\label{eq:lambda_ADMM_update}
\end{equation}

The inner ADMM iterations~\eqref{eq:phi_ADMM_update_closed},~\eqref{eq:z_ADMM_update}, and~\eqref{eq:lambda_ADMM_update} are repeated until the primal and dual residuals
\begin{equation}
\label{rsup}
  \mathbf r^{[r]} \triangleq \boldsymbol\phi^{[r]}-\mathbf z^{[r]}, \mathbf s^{[r]} \triangleq \rho_{\phi}(\mathbf z^{[r]}-\mathbf z^{[r-1]})
\end{equation}
fall below prescribed tolerances. The resulting $\boldsymbol\phi^{[r]}$ is then taken as the updated $\boldsymbol\phi^{(t+1)}(\leftrightarrow\boldsymbol\Phi^{(t+1)})$ in~\eqref{eq:phi_MM_quadratic_final}.
\subsubsection{Update of $\boldsymbol\zeta$}
After updating $\mathbf W^{(t+1)}, \boldsymbol\Phi^{(t+1)}$, and so on, $\boldsymbol\zeta$ is updated by:
\begin{equation}
  \zeta_i^{(t+1)}  =
  \zeta_i^{(t)}
  +
  \frac{1}{\rho_1}
  \Big(
    F_i(\mathbf W^{(t+1)},\boldsymbol\Phi^{(t+1)})
    -
    f_i^{(t+1)}
  \Big)~(i=1,\cdots,6).
\label{eq:zeta_update}
\end{equation}
The overall procedure is presented in Algorithm~\ref{alg11}.
\begin{algorithm}[t]
  \caption{Joint Precoder and RIS Phase Design for RIS-ISAC with $K$-RARs}
  \label{alg11}
  \begin{algorithmic}[1]
    \Require
      $\{\mathbf H_{\mathrm{BR}},\mathbf H_{\mathrm{RU},k},\mathbf H_{\mathrm{BU},k}\}_{k=1}^K, \mathbf h_{d,t},\mathbf h_{r,t}, \epsilon, \rho_1,\rho_{\phi}, P_{\max}$
    \State \textbf{Initialize:}
    $\boldsymbol\phi^{(0)}(\leftrightarrow \boldsymbol\Phi^{(0)})$, $\mathbf W^{(0)}$,
    $\mathbf r^{(0)}$, $\{\mathbf c_k^{(0)}\}_{k=1}^K$,
    $\mathbf f^{(0)}$, $\mathbf J_{\mathrm{aux}}^{(0)}\succ\mathbf 0$,
    $\boldsymbol\zeta^{(0)}$, $t \gets 0$
    \While{not converged}
      \State Update $\mathbf r^{(t+1)}$ and $\{\mathbf c_k^{(t+1)}\}$ using~\eqref{rkmax},~\eqref{ckmax}
      \State Update $\mathbf J_{\mathrm{aux}}^{(t+1)}$ and $\mathbf f^{(t+1)}$ using~\eqref{jfsdp_para}
      \State Update $\mathbf W^{(t+1)}$ by solving~\eqref{eq:w_subProb_final}
      \State Initialize $r\gets 0$, $\boldsymbol\phi^{[0]}=\boldsymbol\phi^{(t)}$, $\mathbf z^{[0]}$, $\boldsymbol\lambda^{[0]}$
      \While{not converged}
        \State Update $\boldsymbol\phi^{[r+1]}$ using~\eqref{eq:phi_ADMM_update_closed}
        \State Update $\mathbf z^{[r+1]}$ using~\eqref{eq:z_ADMM_update}
        \State Update $\boldsymbol\lambda^{[r+1]}$ using~\eqref{eq:lambda_ADMM_update}
        \State $r \gets r+1$
      \EndWhile
      \State $\boldsymbol\phi^{(t+1)} \gets \boldsymbol\phi^{[r]}, \boldsymbol\Phi^{(t+1)}\gets\mathrm{diag}(\boldsymbol\phi^{(t+1)})$
      \State Clear $\{\boldsymbol\phi^{[r]},\mathbf z^{[r]},\boldsymbol\lambda^{[r]}\}$
      \State Update $\boldsymbol\zeta^{(t+1)}$ using~\eqref{eq:zeta_update}
      \State $\rho_1 \gets 0.8\rho_1, t\gets t+1$
    \EndWhile
    \State \Return $\mathbf W^\star=\mathbf W^{(t)}, \boldsymbol\phi^\star=\boldsymbol\phi^{(t)}$
  \end{algorithmic}
\end{algorithm}
\subsection{Computational Complexity Analysis}
\subsubsection{Updates of $\mathbf r$ and $\{\mathbf c_k\}$}
The updates of $\mathbf r$ and $\{\mathbf c_k\}$ are given in closed form by~\eqref{rkmax} and~\eqref{ckmax}. Their cost is dominated by evaluating $\{\mathbf g_{k,j}\}$ and $B_k$ in~\eqref{eq:Bk_def}. Given $\boldsymbol\Phi$, forming $\mathbf H_{\mathrm{eff},k}^{\mathrm{com}}(\boldsymbol\Phi)\mathbf W$ costs $\mathcal O(M N_t K)$ per user, hence $\mathcal O(K^2 M N_t)$ for all users. The phase-alignment by $\mathbf D_{b,k}$ is diagonal and costs $\mathcal O(M)$ per user. Therefore, the overall complexity per outer iteration is $\mathcal O\left(K^2 M N_t\right)$.
\subsubsection{Update of $(\mathbf J_{\mathrm{aux}},\mathbf f)$}
Problem~\eqref{jfsdp_para} has decision variables $\mathbf J_{\mathrm{aux}}\in\mathbb R^{2\times 2}$ and $\mathbf f\in\mathbb C^{6}$, and the LMI $\mathbf M(\mathbf f,\mathbf J_{\mathrm{aux}})\succeq \mathbf 0$ has size $4\times 4$. Hence, the SDP dimension is constant, and the solver cost is $\mathcal O(1)$ per call in big-$\mathcal O$ sense.
\subsubsection{Update of $\mathbf W$}
For the $\mathbf W$-update, we solve the QP in~\eqref{eq:w_subProb_final} over $\mathbf w=\mathrm{vec}(\mathbf W)$. An interior-point method for a convex QP with $2N_tK$ real decision variables has worst-case computational complexity $\mathcal O\big((N_tK)^3\big)$ per outer
iteration~\cite{boyd}. In addition, assembling the QP coefficients in~\eqref{eq:w_subProb_final} incurs extra, but lower-order, overhead. Specifically, this step requires forming $\mathbf B^*\mathbf B$ and the (rank-one) matrices associated with $\mathrm{vec}(\mathbf A_i(\boldsymbol\Phi)\mathbf W^{(t)}) \mathrm{vec}(\mathbf A_i(\boldsymbol\Phi)\mathbf W^{(t)})^*$ (and their conjugate counterparts) for $i=1,\cdots,6$. Computing $\mathbf A_i(\boldsymbol\Phi)$ from~\eqref{eq:Ai_explicit} involves products of matrices that inherit the rank-one (outer-product) structure of $\mathbf H_t(\boldsymbol\Phi,\boldsymbol\theta)$ and its derivatives, so each
$\mathbf A_i(\boldsymbol\Phi)$ can be formed in $\mathcal O(N_t^2)$ flops. Multiplying
$\mathbf A_i(\boldsymbol\Phi)$ with $\mathbf W^{(t)}\in\mathbb C^{N_t\times K}$ then costs $\mathcal O(N_t^2K)$ flops per $i$. Therefore, the overall complexity per outer iteration is $\mathcal O\big((N_tK)^3\big)$.
\subsubsection{Update of $\boldsymbol\phi$}
The update of $\boldsymbol\phi$ can be decomposed into following two stages:
\paragraph*{\textbf{Gradient evaluation}}
The MM surrogate~\eqref{eq:Fi_MM_majorizer} requires $\mathbf d_i^{(t)}=\nabla_{\boldsymbol\phi}F_i(\mathbf W,\boldsymbol\Phi^{(t)})$ for $i=1,\cdots,6$. By~\eqref{eq:dit_component_appendix_final}, the $n$th entry of $\mathbf d_i^{(t)}$ involves $\mathrm{tr}\big(\frac{\partial \mathbf A_i}{\partial \phi_n}\mathbf R_x\big)$.  Substituting the channel derivative expressions in~\eqref{eq:Ht_phi_deriv_appendix},~\eqref{eq:HB_phi_deriv_appendix}, and the definitions of $\mathbf H_t$, $\mathbf H_{\mathrm B}$, and $\mathbf H_{\mathrm R}$ in~\eqref{eq:HR_def_appendix} into~\eqref{aidefphi} and exploiting the cyclic property of the trace, each $\mathrm{tr}\big(\frac{\partial \mathbf A_i}{\partial \phi_n}\mathbf R_x\big)$ reduces to a linear combination of a constant number of scalar inner-products between $\mathbf s_n$ or $\mathbf s'_n$ and a few $N_t$-dimensional vectors such as $\mathbf R_x\mathbf v_t^{*}$, $\mathbf R_x\mathbf v_{\mathrm B}^{*}$, and $\mathbf R_x\mathbf v_{\mathrm R}^{*}$. The latter can be precomputed once, yielding an $\mathcal O(N_t^2)$ cost, while the former requires evaluating all $N$ gradient entries with $\mathcal O(NN_t)$ operations. Hence, the overall complexity per iteration scales as $\mathcal O(N_t^2+NN_t)$.
\paragraph*{\textbf{ADMM iterations}}
Within each $t$,~\eqref{eq:phi_ADMM_update_closed} requires solving a linear system with
$\big(2\widehat{\mathbf D}^{(t)}+\rho_\phi\mathbf I\big)\in\mathbb C^{N\times N}$. By precomputing a Cholesky factorization once per $t$, this step costs $\mathcal O(N^3)$. Each subsequent ADMM iteration solves two triangular systems with complexity $\mathcal O(N^2)$, while~\eqref{eq:z_ADMM_update} and~\eqref{eq:lambda_ADMM_update} are elementwise and cost $\mathcal O(N)$. Hence, the complexity scales with $\mathcal{O}(RN^2)$ with the number of inner $R$ ADMM iterations. Combining these, the total computational complexity per outer iteration is given by $\mathcal O\left(N_t^2 + N N_t + N^3 + R N^2\right)$. 
\subsubsection{Overall complexity}
Collecting the dominant costs, the overall complexity of Algorithm~\ref{alg11} with $T$ outer iterations is
\begin{equation}
\label{compalg}
\mathcal O\left(T\Big[K^2 MN_t+(N_tK)^3+NN_t+N^3+RN^2\Big]\right).
\end{equation}
\section{Simulation Results}
\subsection{Parameter Setup}
We first define the received SNR:
\begin{equation}
\label{resnr}
\mathrm{SNR}\triangleq\frac{\mathbb{E}\left[\big\|\mathbf H_{\mathrm{eff}}^{\mathrm{com}}(\boldsymbol\Phi)\mathbf x\big\|_2^2\right]} {\mathbb{E}\left[\|\mathbf n\|_2^2\right]}.
\end{equation}
Furthermore, to quantify the relative strength of the reference signal, we introduce the reference-to-signal ratio (RSR)~\cite{Precoding_atomicMIMO}:
\begin{equation}
\label{rsrdef}
\mathrm{RSR}\triangleq \frac{\mathbb{E}\left[\|\mathbf b\|_2^2\right]} {\mathbb{E}\left[\big\|\mathbf H_{\mathrm{eff}}^{\mathrm{com}}(\boldsymbol\Phi)\mathbf x+\mathbf n\big\|_2^2\right]},
\end{equation}
which quantifies the relative power level of the reference signal with respect to the received communication signal and noise.
\begin{figure}[t]
    \centering
    \subfloat[]{%
        \includegraphics[width=0.24\textwidth]{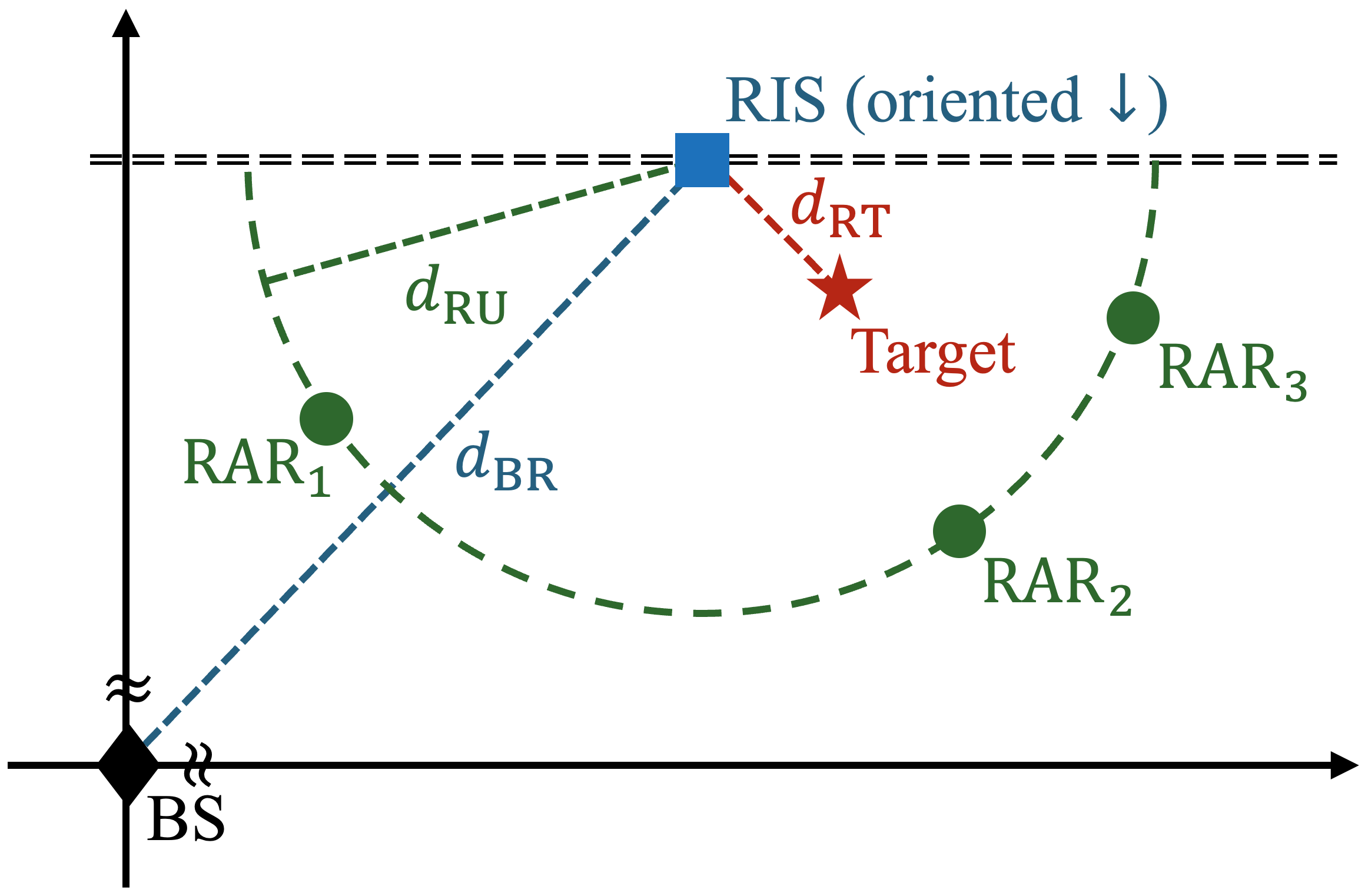}
        \label{fig_sim}%
    }
    \subfloat[]{%
        \includegraphics[width=0.24\textwidth]{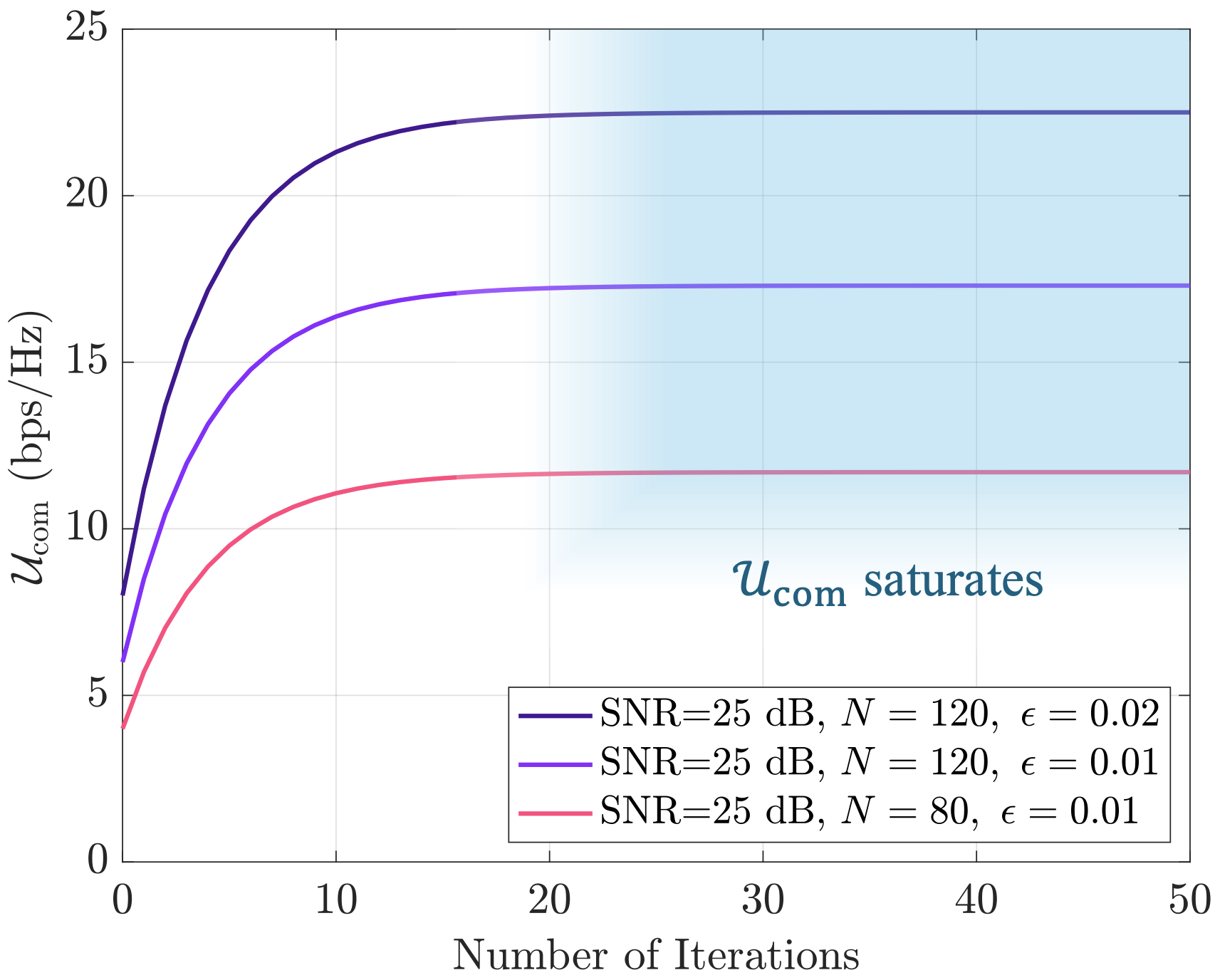}
        \label{fig_iter}%
    }
    \caption{(a) Illustration of the position of BS, RIS, RARs, and target and (b) $\mathcal{U}_{\mathrm{com}}$ versus the number of iterations.}
    \label{fig_0}
\end{figure}

Unless otherwise stated, the simulations are conducted with the parameters in Table~\ref{tabsim}. We set $\varphi_{\mathrm{B},t}=\theta_{\mathrm{R}}=\frac{\pi}{4}$ and $\varphi_{\mathrm{R},t}=-\frac{\pi}{4}$ for the DoD and DoAs, where $\theta_{\mathrm{B}}$ is explicitly given by $\theta_{\mathrm{B}}=\tan^{-1}(\frac{d_{\mathrm{BR}} \sin\varphi_{\mathrm{B},t}-d_{\mathrm{RT}}\cos \theta_{\mathrm{R}}}{d_{\mathrm{BR}} \cos\varphi_{\mathrm{B},t}+d_{\mathrm{RT}}\sin \theta_{\mathrm{R}}})$, and distances of the BS-RIS, the RIS-target, and the RIS-$\forall$user links as $d_{\mathrm{BR}}=50$~m, $d_{\mathrm{RT}}=3$~m, and $d_{\mathrm{RU}}=8$~m, respectively~\cite{Liu_RIS_ISAC_CRB}, where all users are placed within the RIS-facing sector to ensure they remain inside its effective reflection region. Since our focus is the estimation of $\boldsymbol\theta$, we assume $\alpha_t=1$ for simplicity. Fig.~\ref{fig_sim} depicts the simulation environment.

For the atomic configuration, the Rydberg energy levels $52D_{5/2}$ and $53P_{3/2}$ are adopted for detecting the $f=5$~GHz frequency. Utilizing~\cite{rydpar}, $\boldsymbol{\mu}_{\mathrm{eg}}$ over $52D_{5/2}$ and $53P_{3/2}$ is calculated as $[0, 1785.916qa_0, 0]^{\mathrm{T}}$, where $a_0 =5.292\times10^{-11}$~m specifies the Bohr radius, and $q = 1.602 \times10^{-19}$~C is the charge of an electron. Thereafter, $\boldsymbol{\epsilon}_{k,m,b}$ and $\boldsymbol{\epsilon}_{m,n,k,\ell}$ are randomly generated on unit circles orthogonal to their respective incident directions. All channel coefficients are configured according to the same parameter settings in~\cite[Table~I]{atomicjsac}, and Monte Carlo trials are conducted $10^3$ times for every simulation. For the LO-RAR link, since the separation between the LO-RAR is sufficiently small, we approximate $\{\rho_{k,m,b}\}$ by $\rho_{b}$ and model $\forall\phi_{k,m,b}$ as being uniformly distributed over $[0, 2\pi)$. To make $\mathcal{U}_{\mathrm{com}}$ more familiar and easily interpretable, we evaluate in simulations using [bps/Hz] instead of nats.

\begin{table}[t]
\centering
\caption{System Parameters}
\label{tabsim}
\begin{tabular}{l c}
\toprule
\textbf{Parameter} & \textbf{Value} \\
\midrule
Number of RAR-aided users $K$ (unless referred) & 3 \\
Number of RAR elements $M$ (unless referred) & 35 \\
Received SNR in~\eqref{resnr} & 25~dB \\
Number of RIS elements $N$ (unless referred) & 100 \\
RSR in~\eqref{rsrdef} & 10~dB \\
CRB constraint $\epsilon$ (unless referred) & 0.025 \\
Number of radar snapshots $L$ & 1024 \\
Rician $K$-factor $\kappa$ & 2 \\
\bottomrule
\end{tabular}
\end{table}

\subsection{Reliability of the Proposed AO Framework}
Fig.~\ref{fig_iter} shows $\mathcal{U}_{\mathrm{com}}$ as a function of the number of iterations in Algorithm~\ref{alg11}. We observe that the proposed framework converges within 20 iterations, demonstrating fast convergence and confirming the effectiveness of the proposed algorithm.
\subsection{Performance Comparison under Several Effects}
\begin{figure}[t]
    \centering
    \subfloat[]{%
        \includegraphics[width=0.16\textwidth]{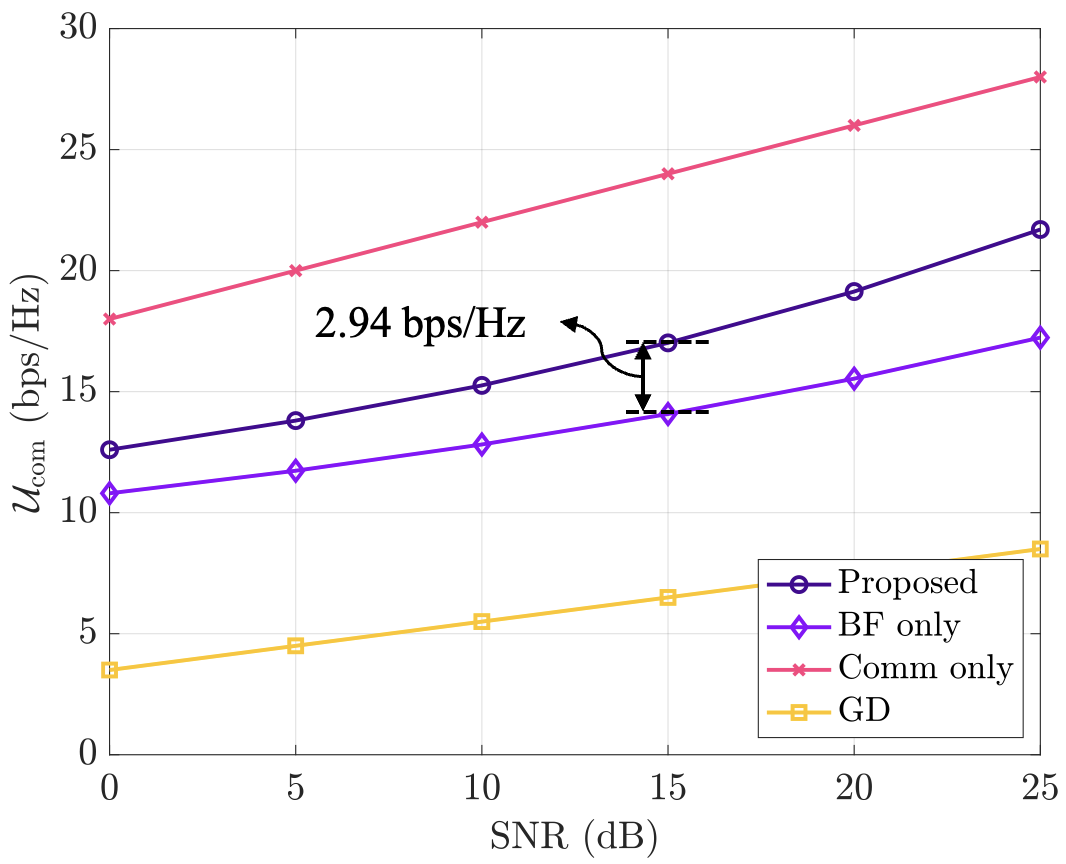}
        \label{fig_snr}%
    }
    \subfloat[]{%
        \includegraphics[width=0.16\textwidth]{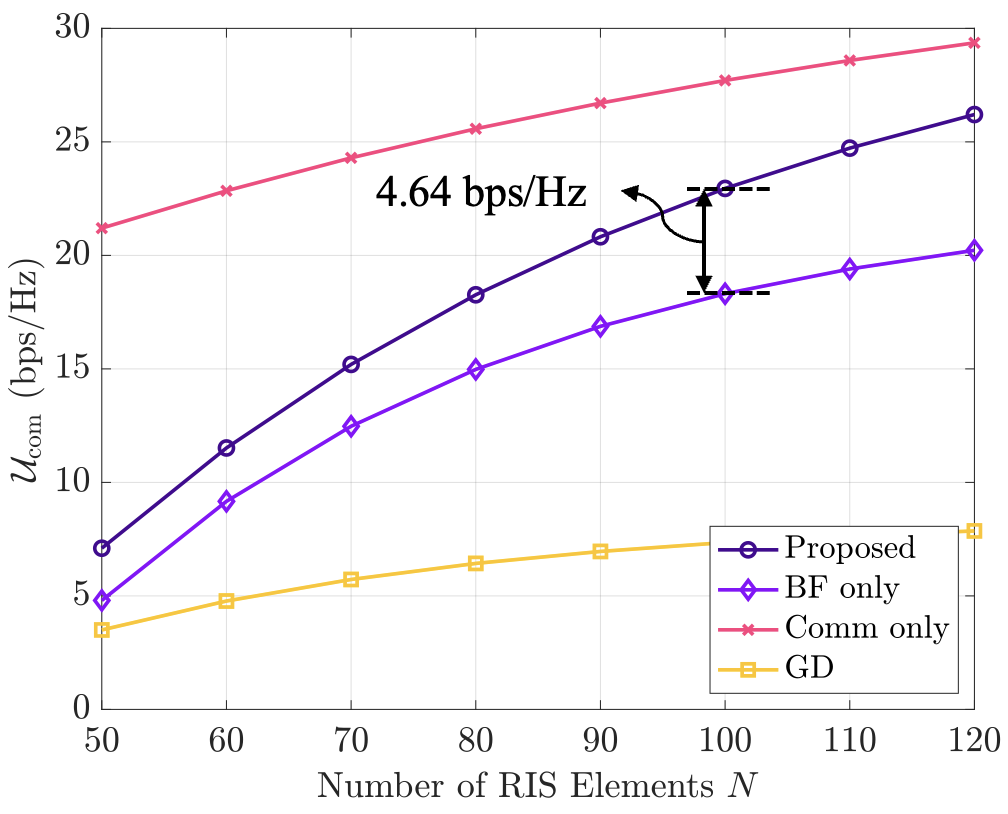}
        \label{fig_ris}%
    }
     \subfloat[]{%
        \includegraphics[width=0.16\textwidth]{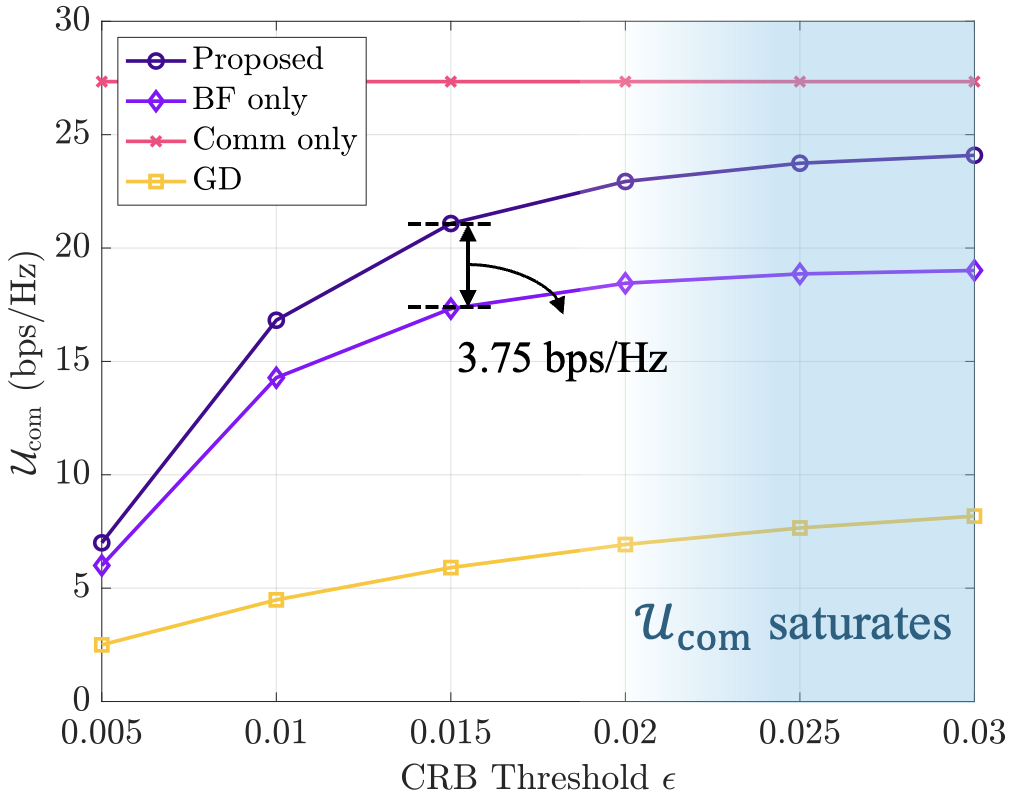}
        \label{fig_crb}%
    }
    \caption{$\mathcal{U}_{\mathrm{com}}$ versus (a) received SNR (b) the number of RIS elements and (c) the CRB constraint.}
    \label{fig_1}
\end{figure}

To demonstrate the performance advantages of the proposed framework (labeled as \textbf{``Proposed''} in the figures), we compare it with the following benchmark schemes:
\begin{itemize}
\item \textbf{``Comm-only''}: Only the communication objective with multi-RAR is optimized, while the sensing constraint is ignored. The proposed optimization framework is employed with the sensing-related constraints removed.
\item \textbf{``BF-only''}: $\boldsymbol\Phi$ is fixed and obtained by maximizing the sum effective BS-RAR channel gain~\cite{Liu_RIS_ISAC_CRB, JAP}. Thereafter, $\mathbf W$ is iteratively optimized using Algorithm~\ref{alg11}.
\item \textbf{``GD''}: $\mathbf W$ and $\boldsymbol\phi$ are alternately updated via projected gradient ascent on a penalized objective with CRB constraint and Riemannian gradient ascent method~\cite{projg, Rieman1}, respectively; After each gradient step, $\mathbf W$ is projected onto the transmit-power constraint in~\eqref{eq:power_constraint} and $\boldsymbol\phi$ is projected onto the unit-modulus set by phase normalization.
\end{itemize}
Fig.~\ref{fig_snr} illustrates $\mathcal{U}_{\mathrm{com}}$ as a function of the received SNR. The proposed framework, which jointly optimizes the RIS configuration and RAR-aided communication beamforming, achieves a significant performance improvement over the ``BF-only" and ``GD" schemes; for example, the performance gap relative to the ``BF-only" scheme is approximately 2.94~bps/Hz at $\mathrm{SNR}=15$~dB. This improvement arises because higher SNR enhances the reliability of RAR-side communication observations, improving the conditioning of the effective received signal. Consequently, more power and DoF can be allocated to communication, yielding a steadily increasing $\mathcal{U}_{\mathrm{com}}$ and a growing performance ratio relative to the ``Comm-only'' scheme. By contrast, the ``BF-only" scheme lacks CRB-aware adaptation: $\boldsymbol\Phi$ is optimized solely for instantaneous channel gain and cannot exploit sensing-communication tradeoffs, resulting in limited gains even at high SNR. Similarly, the ``GD" scheme provides only marginal improvements due to ineffective updates under strong nonconvexity\cite{projg,Rieman1}, a trend that persists across the remaining simulations.

Fig.~\ref{fig_ris} illustrates $\mathcal{U}_{\mathrm{com}}$ as a function of $N$. As $N$ increases, the proposed framework exhibits a pronounced performance improvement compared with the ``BF-only" and ``GD" schemes; for example, the performance gap between the proposed design and the ``BF-only" scheme is approximately 4.64~bps/Hz at $N=100$. This improvement arises because enlarging the RIS aperture enhances spatial focusing capability~\cite{JAP}, thereby reducing the beamforming effort required to satisfy the sensing constraint. In our formulation, the CRB constraint acts as the dominant limiting factor. As $N$ increases, the RIS-induced Fisher Information grows~\cite{Liu_RIS_ISAC_CRB}, which relaxes the CRB constraint. Consequently, the feasible set of $\mathbf W$ expands and the $\boldsymbol\Phi$-update requires less distortion of communication-favorable phase configurations, allowing more power and DoF to be allocated toward enhancing $\mathcal{U}_{\mathrm{com}}$. In contrast, the ``BF-only" scheme optimizes $\boldsymbol\Phi$ solely for channel gain; thus, although $N$ increases, the sensing burden on beamforming is not reduced, preventing RIS aperture growth from translating into comparable communication gains.

\begin{figure}[t]
    \centering
    \subfloat[]{%
        \includegraphics[width=0.16\textwidth]{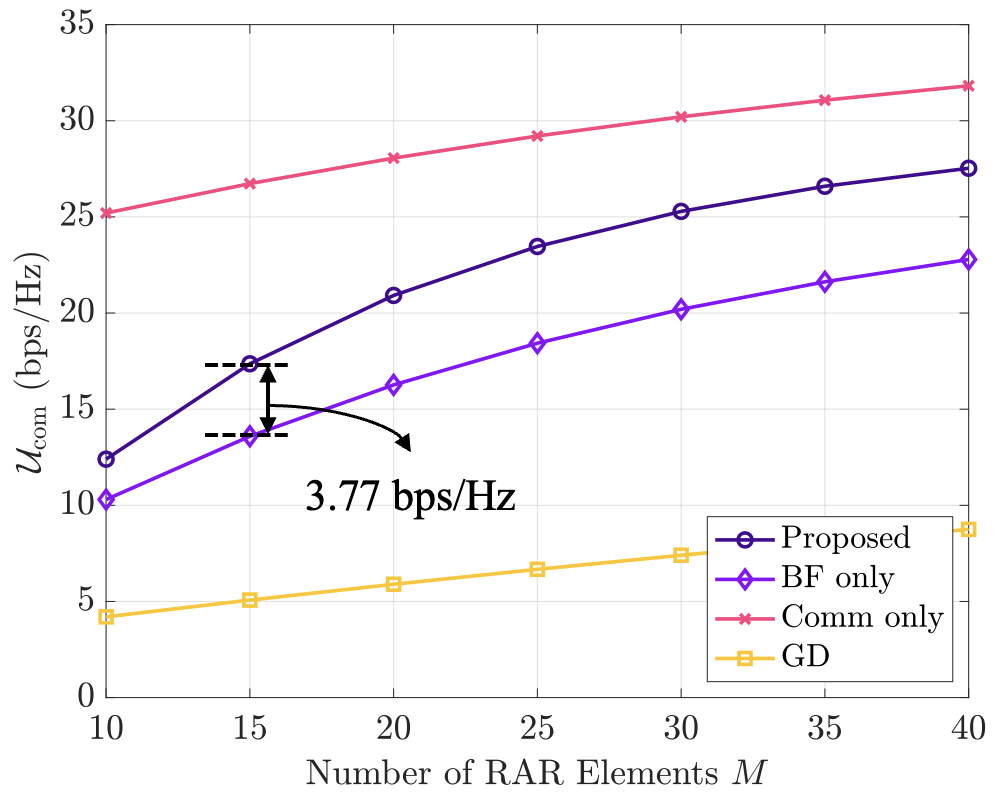}
        \label{fig_rar}%
    }
    \subfloat[]{%
        \includegraphics[width=0.16\textwidth]{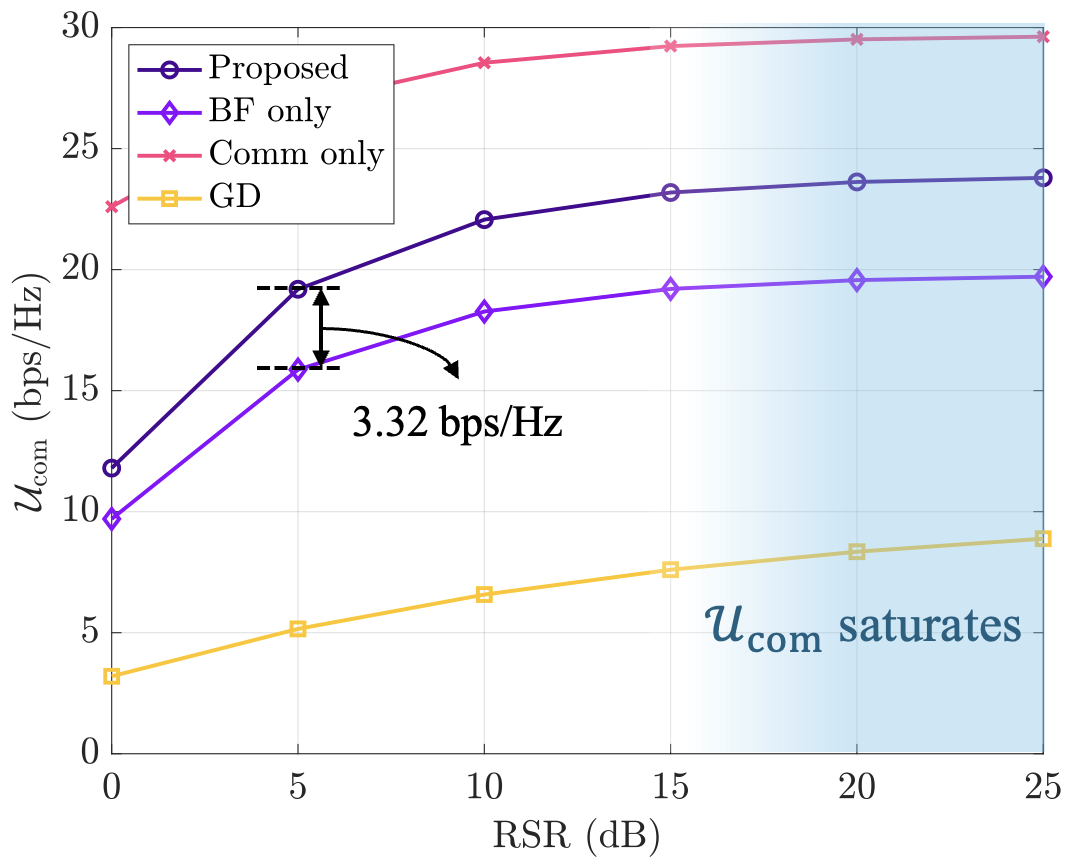}
        \label{fig_rsr}%
    }
     \subfloat[]{%
        \includegraphics[width=0.16\textwidth]{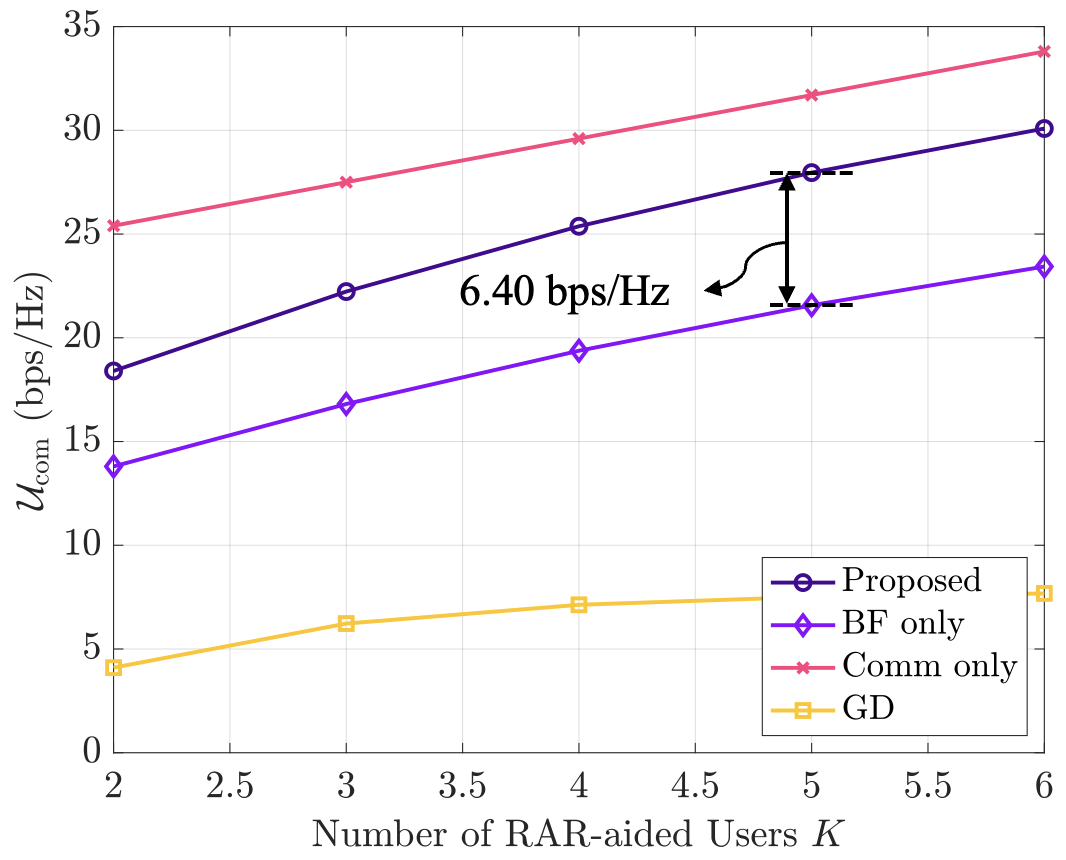}
        \label{fig_rark}%
    }
    \caption{$\mathcal{U}_{\mathrm{com}}$ versus (a) the number of RAR elements (b) RSR and (c) number of RAR-aided users.}
    \label{fig_2}
\end{figure}
Fig.~\ref{fig_crb} illustrates $\mathcal{U}_{\mathrm{com}}$ as a function of $\epsilon$. As $\epsilon$ increases, the sensing accuracy requirement is gradually relaxed, enabling $\mathbf W$ and $\boldsymbol\Phi$ to more effectively exploit the available spatial DoF and transmit power for communication. Herein, the ``Comm-only" scheme remains insensitive to $\epsilon$, since it ignores the sensing constraint and thus serves as a constant upper bound. We further observe that the proposed joint design consistently outperforms the ``BF-only" and ``GD" over the entire range of $\epsilon$; for example, the performance gap between ``Proposed" and ``BF-only" is approximately 3.75~bps/Hz at $\epsilon=0.015$. As $\epsilon$ continues to increase, the CRB constraint becomes progressively less restrictive and eventually inactive, causing the proposed design to approach the ``Comm-only" bound.


Fig.~\ref{fig_rar} illustrates $\mathcal{U}_{\mathrm{com}}$ as a function of $M$. As $M$ increases, the proposed scheme achieves a steadily increasing communication utility and consistently outperforms the ``BF-only" and ``GD" schemes; for example, the performance gap between the proposed design and the ``BF-only" scheme is approximately 3.77~bps/Hz at $M=15$. This improvement arises because a larger $M$ provides the RAR receiver with additional communication observation dimensions, thereby improving the conditioning and spatial diversity of $\{\widetilde{\mathbf E}_k^{\mathrm{com}}(\boldsymbol\Phi)\}$. This effect indirectly relaxes the CRB constraint in the optimization, as less beamforming effort is required to maintain reliable communication-side signal extraction. Consequently, more power and DoF can be allocated to communication, yielding the observed gain in $\mathcal{U}_{\mathrm{com}}$. By contrast, the ``BF-only" scheme optimizes $\boldsymbol\Phi$ solely for channel-gain maximization and ignores the sensing constraint. As a result, although increasing $M$ offers additional communication-side observations, this information is not incorporated into the optimization, and enlarging the RAR array does not translate into substantial communication gains.

Fig.~\ref{fig_rsr} illustrates $\mathcal{U}_{\mathrm{com}}$ as a function of RSR. Herein, a higher RSR strengthens the reference illumination used in RAR-based symbol extraction, thereby improving the stability of communication-side observations. Consequently, the proposed framework can satisfy the sensing requirement using a smaller fraction of transmit power, allowing more DoF and power to be allocated toward enhancing the communication rate. This yields a notable performance gain; for example, an improvement of approximately 3.32~bps/Hz over the ``BF-only" scheme at $\mathrm{RSR}=5$~dB. In contrast, the ``BF-only" scheme does not incorporate sensing information into its design: $\boldsymbol\Phi$ is fixed solely to maximize channel gain and does not adapt to the reference quality. As a result, increasing RSR does not alleviate the communication-sensing coupling in BF-only, and additional reference strength yields only marginal gains. When RSR becomes sufficiently large, all schemes gradually saturate because the sensing accuracy requirement is already satisfied (i.e., the CRB is sufficiently small) and no longer constitutes the dominant bottleneck; hence further increases in LO power yield diminishing returns.

Fig.~\ref{fig_rark} illustrates $\mathcal{U}_{\mathrm{com}}$ as a function of $K$. The proposed joint design consistently outperforms the ``BF-only" and ``GD" schemes over the entire range of $K$; for example, it achieves a performance gain of about 6.40~bps/Hz over the ``BF-only"' scheme when $K=5$. This behavior arises because, as $K$ increases, the achievable rate $\mathcal{U}_{\mathrm{com}}$ becomes increasingly constrained by multi-user interference and finite transmission resources~\cite{summimo, jindal, CBF22}. In this regime, interference mitigation is inherently embedded in the algorithm: (i) $\{r_k\}$ is updated based on SINR expressions whose denominators include interference terms, thereby penalizing interference growth; (ii) during the $\mathbf W$-update, $\{\|\mathbf g_{k,j}\|^2\}_{j\neq k}$ explicitly appear in the gradient, steering the solution away from interference-dominant directions; and (iii) in the $\boldsymbol\Phi$-update, the RAR-induced response adapts power distribution and path reinforcement, strengthening desired links and indirectly reducing interference impact. By contrast, the ``BF-only" employs a separated design in which $\boldsymbol\Phi$ is optimized without accounting for the sensing constraint, resulting in a persistent sensing-induced beamforming cost and a widening performance gap.


\section{Conclusion}
In this paper, we proposed an RIS-ISAC framework for multi-user systems employing RARs. By explicitly incorporating sensing accuracy requirements into the joint design of the transmit beamformer and RIS configuration, the proposed approach satisfies the CRB constraint with reduced effective beamforming effort, thereby enhancing the communication utility in terms of sum-rate. An efficient joint optimization algorithm was developed to address the inherent nonconvexity, enabling a systematic BCD framework through a sequence of tractable subproblems based on FP, MM, and ADMM updates. Extensive simulation results demonstrated that the proposed design consistently outperforms conventional schemes across diverse operating conditions and exhibits a progressively smaller performance gap relative to the communication-only benchmark as system resources increase. Overall, these results confirm that tight integration of RAR and RIS-ISAC constitutes a viable and effective design paradigm for fully exploiting the potential of RARs in 6G.


\section*{Appendix A}
\section*{Explicit Form of $\mathbf d_i^{(t)}$}
For convenience, define the RIS-BS coupling matrix $\mathbf S
\triangleq
\mathbf H_{\mathrm {BR}}^{\mathrm T}
\mathrm{diag}(\mathbf h_{r,t})
\in\mathbb C^{N_t\times N}$ and its $n$th column $\mathbf s_n \triangleq \mathbf S(:,n)$, so that $\mathbf v_t(\boldsymbol\Phi)$ can be written as $\mathbf v_t(\boldsymbol\Phi)=\mathbf h_{d,t} + \mathbf S\boldsymbol\phi, \frac{\partial \mathbf v_t}{\partial \phi_n} = \mathbf s_n$. The derivative of $\mathbf H_t=\mathbf v_t\mathbf v_t^{\mathrm T}$ with respect to the $\phi_n$ is
\begin{equation}
\label{eq:Ht_phi_deriv_appendix}
\frac{\partial\mathbf H_t}{\partial\phi_n}
=
\mathbf s_n \mathbf v_t^{\mathrm T}
+
\mathbf v_t \mathbf s_n^{\mathrm T}.
\end{equation}
For differentiating $\mathbf H_t$ with respect to $\theta_{\mathrm{B}}$ and $\theta_{\mathrm{R}}$, let $\mathbf v_{\mathrm B}
\triangleq
\frac{\partial\mathbf v_t}{\partial\theta_{\mathrm B}}$ and $\mathbf v_{\mathrm R}
\triangleq\frac{\partial\mathbf v_t}{\partial\theta_{\mathrm R}}$. Then by definition
\begin{equation}
\mathbf H_{\mathrm B}
\triangleq
\frac{\partial\mathbf H_t}{\partial\theta_{\mathrm B}}
=
\mathbf v_{\mathrm B}\mathbf v_t^{\mathrm T}
+
\mathbf v_t\mathbf v_{\mathrm B}^{\mathrm T},
\mathbf H_{\mathrm R}
\triangleq
\frac{\partial\mathbf H_t}{\partial\theta_{\mathrm R}}
=
\mathbf v_{\mathrm R}\mathbf v_t^{\mathrm T}
+
\mathbf v_t\mathbf v_{\mathrm R}^{\mathrm T}.
\label{eq:HR_def_appendix}
\end{equation}
Since $\mathbf h_{d,t}$ does not depend on $\boldsymbol\phi$, so is $\mathbf v_{\mathrm B}$ and hence $\frac{\partial\mathbf v_{\mathrm B}}{\partial\phi_n}=\mathbf 0$. On the other hand, $\mathbf v_{\mathrm R}$ depends on $\boldsymbol\phi$
through the RIS-assisted link. Using the chain rule, $\mathbf v_{\mathrm R}=\mathbf H_{\mathrm {BR}}^{\mathrm T}\boldsymbol\Phi\frac{\partial\mathbf h_{r,t}}{\partial\theta_{\mathrm R}}$, so that we can introduce $\mathbf S'
\triangleq
\mathbf H_{\mathrm {BR}}^{\mathrm T}
\mathrm{diag}\left(
\frac{\partial\mathbf h_{r,t}}{\partial\theta_{\mathrm R}}
\right)$ and its $n$th column $\mathbf s'_n \triangleq \mathbf S'(:,n)$, and write $\mathbf v_{\mathrm R}=\mathbf S'\boldsymbol\phi, \frac{\partial\mathbf v_{\mathrm R}}{\partial\phi_n}=\mathbf s'_n$. Hence, we obtain
\begin{equation}
\begin{aligned}
\label{eq:HB_phi_deriv_appendix}
\frac{\partial\mathbf H_{\mathrm B}}{\partial\phi_n}
&=
\mathbf v_{\mathrm B}\mathbf s_n^{\mathrm T}
+
\mathbf s_n\mathbf v_{\mathrm B}^{\mathrm T},
\\
\frac{\partial\mathbf H_{\mathrm R}}{\partial\phi_n}
&=
\mathbf s'_n \mathbf v_t^{\mathrm T}
+
\mathbf v_{\mathrm R}\mathbf s_n^{\mathrm T}
+
\mathbf s_n \mathbf v_{\mathrm R}^{\mathrm T}
+
\mathbf v_t \mathbf s_n^{'\mathrm T}.
\end{aligned}
\end{equation}
Directly differentiating~\eqref{eq:Ai_explicit} with respect to $\phi_n$ yields~\eqref{aidefphi}, 
\begin{figure*}
\begin{equation}
\begin{aligned}
\frac{\partial\mathbf A_1}{\partial\phi_n}
&=
L\left(
\frac{\partial\mathbf H_{\mathrm B}^{*}}{\partial\phi_n}\mathbf H_{\mathrm B}
+
\mathbf H_{\mathrm B}^{*}\frac{\partial\mathbf H_{\mathrm B}}{\partial\phi_n}
\right), 
\frac{\partial\mathbf A_2}{\partial\phi_n}=
L\left(
\frac{\partial\mathbf H_{\mathrm B}^{*}}{\partial\phi_n}\mathbf H_{\mathrm R}
+
\mathbf H_{\mathrm B}^{*}\frac{\partial\mathbf H_{\mathrm R}}{\partial\phi_n}
\right), 
\frac{\partial\mathbf A_3}{\partial\phi_n}=
L\alpha_t^{*}\left(
\frac{\partial\mathbf H_{\mathrm B}^{*}}{\partial\phi_n}\mathbf H_t
+
\mathbf H_{\mathrm B}^{*}\frac{\partial\mathbf H_t}{\partial\phi_n}
\right),
\\
\frac{\partial\mathbf A_4}{\partial\phi_n}
&=
L\left(
\frac{\partial\mathbf H_{\mathrm R}^{*}}{\partial\phi_n}\mathbf H_{\mathrm R}
+
\mathbf H_{\mathrm R}^{*}\frac{\partial\mathbf H_{\mathrm R}}{\partial\phi_n}
\right), \frac{\partial\mathbf A_5}{\partial\phi_n}
=
L\alpha_t^{*}\left(
\frac{\partial\mathbf H_{\mathrm R}^{*}}{\partial\phi_n}\mathbf H_t
+
\mathbf H_{\mathrm R}^{*}\frac{\partial\mathbf H_t}{\partial\phi_n}
\right), 
\frac{\partial\mathbf A_6}{\partial\phi_n}
=
L\left(
\frac{\partial\mathbf H_t^{*}}{\partial\phi_n}\mathbf H_t
+
\mathbf H_t^{*}\frac{\partial\mathbf H_t}{\partial\phi_n}
\right).
\label{aidefphi}
\end{aligned}
\end{equation}
\hrule
\end{figure*}
and along with~\eqref{eq:Fi_quad_form}, the $n$th element of $\mathbf d_i^{(t)}$ is
\begin{equation}
\label{eq:dit_component_appendix_final}
\big[\mathbf d_i^{(t)}\big]_n
=
\mathrm{tr}\left(
\frac{\partial\mathbf A_i(\boldsymbol\Phi)}{\partial\phi_n}
\mathbf R_x
\right)
\Bigg|_{\boldsymbol\Phi=\boldsymbol\Phi^{(t)}},
\end{equation}
and collecting all entries yields the explicit form of $\mathbf d_i^{(t)}$.
\bibliographystyle{IEEEtran}
\bibliography{IEEEexample}

@string{cup="Cambridge University Press"}

@string{oup="Oxford University Press"}

@string{prh="Prentice Hall"}

@string{CONF_ICC			= "Proc. IEEE Int. Conf. on Commun. (ICC)"}

@string{IEEE_J_COMSURVTUT	= "{IEEE} Commun. Surveys Tuts."}

@string{IEEE_J_JSAC			= "{IEEE} J. Sel. Areas Commun."}

@string{IEEE_J_COM			= "{IEEE} Trans. Commun."}

@string{IEEE_J_IT			= "{IEEE} Trans. Inf. Theory"}

@string{IEEE_J_STSP			= "{IEEE} J. Sel. Topics Signal Process."}

@string{IEEE_J_TSP			= "{IEEE} Trans. Signal Process."}

@string{IEEE_J_TVT			= "{IEEE} Trans. Veh. Technol."}

@string{IEEE_J_WCOM			= "{IEEE} Trans. Wireless Commun."}

@ARTICLE{LingRIS,
  author={Dai, Linglong and others},
  	journal={IEEE Access}, 
	title={Reconfigurable Intelligent Surface-Based Wireless Communications: Antenna Design, Prototyping, and Experimental Results}, 
	year={2020},
	volume={8},
	number={},
	pages={45913-45923},
}

@ARTICLE{JAP,
	author={Q. {Wu} and R. {Zhang}},
	journal=IEEE_J_WCOM, 
	title={Intelligent Reflecting Surface Enhanced Wireless Network via Joint Active and Passive Beamforming}, 
	year={2019},
	volume={18},
	number={11},
	month=Nov,
	pages={5394-5409},
}

@book{boyd,
	title={Convex Optimization},
	author={Boyd, S and Vandenberghe, L},
	year={2004},
	publisher={Cambridge, UK: Cambridge Univ. Press},
}

@ARTICLE{RIST,
  author={Wu, Qingqing and others},
   journal={IEEE Trans. Commun.}, 
  title={Intelligent Reflecting Surface-Aided Wireless Communications: A Tutorial}, 
  year={2021},
  volume={69},
  number={5},
  pages={3313-3351},
  month=may,
  }

@ARTICLE{CBF22,
  author={C.-B. {Chae} and others},
  journal=IEEE_J_TSP,
  title={Coordinated beamforming for the Multiuser {MIMO} Broadcast Channel With Limited Feedforward}, 
  year={2008},
  volume={56},
  number={12},
  pages={6044-6056},
  month=dec,
  }

@ARTICLE{Rieman1,
  author={AlaaEldin, Mahmoud and others},
  journal={IEEE Internet Things J.},
  title={Optimization of Energy-Constrained {IRS-NOMA} Using a Complex Circle Manifold Approach}, 
  year={2024},
  volume={11},
  number={20},
  pages={33133-33150},
  month=oct,
  }

@ARTICLE{fracp,
  author={Shen, Kaiming and Yu, Wei},
  journal=IEEE_J_TSP, 
  title={Fractional Programming for Communication Systems—{Part I}: Power Control and Beamforming}, 
  year={2018},
  volume={66},
  number={10},
  pages={2616-2630},
month=may,
  }

@article{RIS_atomic_MIMO,
	title={{RIS}-assisted Atomic {MIMO} Receiver},
	author={Qihao Peng and others},
	journal={arXiv:2510.15763},
	year={2025}
}

@article{Precoding_atomicMIMO,
	title={{MIMO} Precoding for {Rydberg} Atomic Receivers},
	author={Mingyao Cui and others},
	journal={arXiv:2408.14366v2},
	year={2024}
}

@ARTICLE{Song_CRB_IRS,
  author={Song, Xianxin and others},
  journal=IEEE_J_STSP, 
  title={Intelligent Reflecting Surface Enabled Sensing: Cramér-Rao Bound Optimization}, 
  year={2023},
  volume={71},
  number={},
  pages={2011-2026},
  month=may,
  }

@ARTICLE{Liu_RIS_ISAC_CRB,
  author={Liu, Rang and others},
  journal=IEEE_J_WCOM,
   title={{SNR/CRB}-Constrained Joint Beamforming and Reflection Designs for {RIS-ISAC} Systems}, 
  year={2024},
  volume={23},
  number={7},
  pages={7456-7470},
  month=jul,
  }

@book{Kay,
  title={Fundamentals of statistical signal processing: estimation theory},
  author={Steven M. Kay},
  year={1993},
  publisher=prh,
}

@ARTICLE{atomicjsac,
  author={Cui, Mingyao and others},
  journal=IEEE_J_JSAC, 
  title={Towards Atomic {MIMO} Receivers}, 
  year={2025},
  volume={43},
  number={3},
  pages={659-673},
  month=mar,
  }

@article{rydpar,
title = {{ARC} 3.0: An expanded {Python} toolbox for atomic physics calculations},
journal = {Computer Physics Commun.},
volume = {261},
pages = {107814},
year = {2021},
author = {E.J. Robertson and others},
month=apr,
}

@book{fox,
  title={Quantum Optics: An Introductiony},
  author={M. Fox},
  year={2006},
  publisher=oup,
}

@ARTICLE{summimo,
  author={Louie, Raymond H.Y. and others},
  journal=IEEE_J_COM,
  title={Maximum sum-rate of {MIMO} multiuser scheduling with linear receivers}, 
  year={2009},
  volume={57},
  number={11},
  pages={3500-3510},
  month=nov,
  }

@ARTICLE{jindal,
  author={Jindal, N.},
  journal=IEEE_J_IT, 
  title={{MIMO} Broadcast Channels With Finite-Rate Feedback}, 
  year={2006},
  volume={52},
  number={11},
  pages={5045-5060},
  month=nov,
  }

@ARTICLE{projg,
  author={Hu, Guojie and others},
  journal={IEEE Commun. Lett.}, 
  title={Fluid Antennas-Enabled Multiuser Uplink: A Low-Complexity Gradient Descent for Total Transmit Power Minimization}, 
  year={2024},
  volume={28},
  number={3},
  pages={602-606},
  month=mar,
  }

@ARTICLE{efm,
  author={Liu, Bang and others},
  journal={Electromagn. Sci.}, 
  title={Electric Field Measurement and Application Based on {Rydberg} Atoms}, 
  year={2023},
  volume={1},
  number={2},
  pages={1-16},
  month=jun,
  }

@ARTICLE{qmusic,
  author={Kim, Hanvit and others},
  journal={IEEE Wireless Commun. Lett.}, 
  title={Quantum-{MUSIC}: Multiple Signal Classification for Quantum Wireless Sensing}, 
  year={2025},
  volume={14},
  number={6},
  pages={1623-1627},
  month=jun,
  }

@ARTICLE{isacmag,
  author={Chen, Yilong and others},
  journal={IEEE Commun. Mag.}, 
  title={Integrated Sensing, Communication, and Powering: Toward Multi-Functional {6G} Wireless Networks}, 
  year={2025},
  volume={63},
  number={8},
  pages={146-153},
  month=aug,
  }

@ARTICLE{isactut,
  author={Zhang, J. Andrew and others},
  journal=IEEE_J_COMSURVTUT, 
  title={Enabling Joint Communication and Radar Sensing in Mobile Networks—A Survey}, 
  year={2022},
  volume={24},
  number={1},
  pages={306-345},
  month={First quarter}, 
  }

@ARTICLE{isacjsac,
  author={Liu, Fan and others},
  journal=IEEE_J_JSAC, 
  title={Integrated Sensing and Communications: Toward Dual-Functional Wireless Networks for {6G} and Beyond}, 
  year={2022},
  volume={40},
  number={6},
  pages={1728-1767},
  month=jun,
  }

@ARTICLE{isaccsm,
  author={Kaushik, Aryan and others},
  journal={IEEE Commun. Stand. Mag.}, 
  title={Toward Integrated Sensing and Communications for {6G}: Key Enabling Technologies, Standardization, and Challenges}, 
  year={2024},
  volume={8},
  number={2},
  pages={52-59},
  month=jun,
  }

@ARTICLE{risisacoj,
  author={Magbool, Ahmed and others},
  journal={IEEE Open J. Commun. Soc.}, 
  title={A Survey on Integrated Sensing and Communication With Intelligent Metasurfaces: Trends, Challenges, and Opportunities}, 
  year={2025},
  volume={6},
  number={},
  pages={7270-7318},
  month=aug,
    }

@ARTICLE{risisacmag22,
  author={Liu, Rang and others},
  journal={IEEE Wireless Commun.}, 
  title={Integrated Sensing and Communication with Reconfigurable Intelligent Surfaces: Opportunities, Applications, and Future Directions}, 
  year={2023},
  volume={30},
  number={1},
  pages={50-57},
  month=feb,
  }

@ARTICLE{Luo_TVT23_RIS_ISAC,
  author={Luo, Honghao and others},
  journal=IEEE_J_TVT, 
  title={{RIS}-Aided Integrated Sensing and Communication: Joint Beamforming and Reflection Design}, 
  year={2023},
  volume={72},
  number={7},
  pages={9626-9630},
  month=jul,
  }

@ARTICLE{Chu_TVT24_Secure_RIS_ISAC,
  author={Chu, Jinjin and others},
  journal=IEEE_J_TVT, 
  title={Joint Beamforming and Reflection Design for Secure {RIS-ISAC} Systems}, 
  year={2024},
  volume={73},
  number={3},
  pages={4471-4475},
  month=mar,
  }

@ARTICLE{IoT23_RIS_Backscatter_ISAC,
  author={Wang, Xinyi and others},
  journal={IEEE Internet Things J.}, 
  title={Integrated Sensing and Communication for {RIS}-Assisted Backscatter Systems}, 
  year={2023},
  volume={10},
  number={15},
  pages={13716-13726},
  month=aug,
  }

@ARTICLE{TWC24_Cooperative_RIS_ISAC,
  author={Yang, Xiaoyu and others},
  journal=IEEE_J_WCOM, 
  title={{RIS}-Assisted Cooperative Multicell {ISAC} Systems: A Multi-User and Multi-Target Case}, 
  year={2024},
  volume={23},
  number={8},
  pages={8683-8699},
  month=aug,
  }

@ARTICLE{IoT24_CRB_mmWave_RIS_ISAC,
  author={Lyu, Wanting and others},
  journal={IEEE Internet Things J.}, 
  title={{CRB} Minimization for {RIS}-Aided {mmWave} Integrated Sensing and Communications}, 
  year={2024},
  volume={11},
  number={10},
  pages={18381-18393},
  month=may,
  }

@ARTICLE{TWC25_MovableAntenna_RIS_ISAC,
  author={Ma, Yaodong and others},
  journal=IEEE_J_WCOM,
  title={Movable-Antenna Aided Secure Transmission for {RIS-ISAC} Systems}, 
  year={2025},
  volume={24},
  number={12},
  pages={10019-10035},
  month=dec,
  }

@ARTICLE{atomicmag,
  author={Cui, Mingyao and others},
  journal={IEEE Commun. Mag.}, 
  title={{Rydberg} Atomic Receiver: Next Frontier of Wireless Communications}, 
  year={2025},
  volume={},
  number={},
  pages={1-7},
}

@ARTICLE{wsat,
  author={Kim, Hanvit and others},
  journal={IEEE Commun. Lett.}, 
  title={Multi-Band Quantum Wireless Sensing for {Rydberg} Atomic Receivers}, 
  year={2025},
  volume={29},
  number={6},
  pages={1476-1480},
  month=jun,
  }

@ARTICLE{TCOM25_Single_RAR_AoA,
  author={Guo, Yuqing and others},
  journal=IEEE_J_COM,
  title={{AoA} Detection Using a Single {Rydberg} Atomic Receiver: Leveraging Inner-Vapor Interference}, 
  year={2025},
  volume={73},
  number={12},
  pages={14828-14844},
  month=dec,
  }

@article{Harnessing_RAR_Tutorial,
	title={Harnessing {Rydberg} Atomic Receivers: From Quantum Physics to Wireless Communications},
	author={Yuanbin Chen and others},
	journal={arXiv:2501.11842v2},
	year={2025}
}

@INPROCEEDINGS{RAR_MU_MIMO_Uplink,
  author={Gong, Tierui and others},
  booktitle=CONF_ICC,
  title={{Rydberg} Atomic Quantum Receivers for the Multi-User {MIMO} Uplink}, 
  year={2025},
  volume={},
  number={},
  pages={4786-4791},
  month=jun,
  }

@article{RAQ_MIMO_Multiband,
	title={{RAQ-MIMO}: {MIMO} for Multi-Band {Rydberg} Atomic Quantum Receiver},
	author={Jieao Zhu and Linglong Dai},
	journal={arXiv:2509.07832},
	year={2025}
}

@article{New_Paradigm_RAR_ISAC,
	title={New Paradigm for Integrated Sensing and Communication with {Rydberg} Atomic Receiver},
	author={Minze Chen and others},
	journal={arXiv:2506.13304v4},
	year={2025}
}

@ARTICLE{RAR_Classical_Comm_Sensing,
  author={Gong, Tierui and others},
  journal={IEEE Wireless Commun.}, 
  title={{Rydberg} Atomic Quantum Receivers for Classical Wireless Communication and Sensing}, 
  year={2025},
  volume={32},
  number={5},
  pages={90-100},
  month=oct,
  }

@ARTICLE{kwakfd,
  author={Kwak, Jong Woo and others},
  journal=IEEE_J_WCOM, 
  title={Analog Self-Interference Cancellation With Practical {RF} Components for Full-Duplex Radios}, 
  year={2023},
  volume={22},
  number={7},
  pages={4552-4564},
  month=jul,
  }

@inproceedings{quanmobi,
author = {Zhang, Fusang and others},
  booktitle={Proc. of the ACM MobiCom},
title = {Quantum Wireless Sensing: Principle, Design and Implementation},
year = {2023},
volume={},
number={},
pages={1-15},
month=oct,
}

@article{qsens,
year = {2015},
month = sep,
volume = {48},
number = {20},
pages = {202001},
author = {Fan, Haoquan and others},
title = {Atom based {RF} electric field sensing},
journal = {J. Phys. B: At. Mol. Opt. Phys.},
}

@ARTICLE{antsp,
  author={Xiao, Zhenyu and others},
  journal=IEEE_J_JSAC, 
  title={Antenna Array Enabled Space/Air/Ground Communications and Networking for {6G}}, 
  year={2022},
  volume={40},
  number={10},
  pages={2773-2804},
  month=oct,
  }

@article{qira,
  title = {Quantum information with {Rydberg} atoms},
  author = {Saffman, M. and others},
  journal = {Rev. Mod. Phys.},
  volume = {82},
  issue = {3},
  pages = {2313-2363},
  year = {2010},
  month =aug,
}

@ARTICLE{tqe,
  author={Fancher, Charles T. and others},
  journal={IEEE Trans. Quantum Eng.}, 
  title={{Rydberg} Atom Electric Field Sensors for Communications and Sensing}, 
  year={2021},
  volume={2},
  number={},
  pages={1-13},
  month=mar,
  }

@article{rarclose3,
  author    = {M. Jing and others},
  title     = {Atomic Superheterodyne Receiver Based on Microwave-Dressed {Rydberg} Spectroscopy},
  journal   = {Nat. Phys.,},
  volume    = {1},
  pages     = {911-915},
  month     = jun,
  year      = {2020}
}

@book{matrix,
  title={Matrix Analysis},
  author={Roger A. Horn and Charles R. Johnson},
  year={2012},
  publisher=cup,
}

\end{document}